# THE EFFECT OF HOUSEWIFE' LABOR ON GDP CALCULATIONS
## Saadet Yağmur KUMCU

## ABSTRACT

In this study, based on the Labor Theory of Value, it has been tried to reveal the evolutionary definitions of "labor" and the developments in the theory, based on the theoretical analysis. Using the example of GDP, which is an indicator of the reflections of labor in the field of social welfare, it has been tried to measure the economic value of housewives' labor through an empirical modelling. For this purpose, in this study conducted with a holistic approach; first of all, the concept of "labor" was questioned in orthodox (mainstream) economic theories; then, by abstracting from the labor-employment relationship, it has been focused on how the unpaid housewife's labor, which is out of employment in the capitalist system, can have an effect on GDP. As an indicator of the need for heterodox policies in the solution of economic problems with the theoretical analysis that is the focus of the study, it was concluded that the changing human profile moved away from rationality and formed the "limited rationality" and accordingly the "heterogeneous individual" profile. The best example of the new heterogeneous individual model, that prefers to make decisions according to the conditions in which needs are met, is considered to be women. Women who make up half of the population as employed and domestic workers; due to their preference to be housewives were defined as the segment that best fits the definition of a limited rational individual, compared to the entire population. Due to these changes in the labor structure and individuals determined as a result of the theoretical analysis, housewife labor was taken into account as the main variable of the empirical analysis of the study. In the first stage of the empirical analysis made with TUIK employment data in the case of Turkey, the inequality measure of the current situation in which housewife labor is not included in GDP was calculated using the Atkinson Inequality Scale. In the second stage of the model, the effect on GDP was calculated by taking into account the labor of housewives. The results of the theoretical and empirical analysis were evaluated in the context of the effect of labor-employment dependency on social inequality and economic welfare. It is hoped that the study will contribute to the literature on heterodox labor theories, which has limited empirical studies.

**CONTENTS**

# LIST OF TABLES

# INDEX OF SYMBOLS AND ABBREVIATIONS

| | |
|---|---|
| EU | European Union |
| USA | United States of America |
| Translator | Translator |
| Ed. | Editor |
| GDP | Gross Domestic Product |
| ILO | International Labor Organization |
| TurkStat | Turkish Statistical Institute |
| UN | United Nations |
| UNDP | United Nations Development Program |

# INTRODUCTION

The main purpose of this research is to define and measure labor, which has been the most fundamental value of humanity since its existence, in order to solve economic problems or to become a society that has reached a level of prosperity. The thesis "The Effect of Housewives' Labor on GDP Calculations"aims to redefine labor with an approach that suggests that the labor of housewives, who are limited rational individuals, in producing goods and services to meet the needs of the entire household should be considered as an economic value.

Labor is the sum of all material and immaterial efforts expended to produce a work. An individual uses his/her labor power to meet the needs of himself/herself and his/her family members or to do a job for a wage. The separation of the work produced by and for whom this labor power is used is considered to be a result of the competitive approach of the capitalist economy. So much so that the labor for cooking, cleaning, ironing and similar tasks, which are usually done by women at home and which are necessary to ensure the continuity of a family member who is "employed" in working life and commutes to and from work, is not accepted as an economic value. In the agricultural period, which began with the transition from nomadic to settled life, agricultural work, which was integrated with domestic life, caused women's labor to become invisible or to be considered so individual as to be ignored in economic terms. While the Renaissance and Reform movements, the French Revolution of 1789 and the Industrial Revolution that followed shaped art, social life and science, it can be observed that women were condemned to domestic work, to be cheap workers in mechanized production or to be the part deprived of education by migrating from villages to cities. With the Industrial Revolution, when the concepts of employer emerged, labor gained a professional meaning and turned into a value for which the owner is paid a wage, while the labor of women in home production is neglected, ignored, and turned into an invisible labor that does not have an economic meaning. With her invisible labor, the woman is also characterized as the hidden actor of injustice within the economic fiction of capitalism (Kumcu, 2019).

In the production model of the capitalist system, labor is considered as a hired force in order to earn wage income and provide some social rights in return. The size of the amount of production, wage income and expenditures made with labor power is accepted as an indicator of the economic development and size of that country as GDP value.

According to Paul A. Samuelson, GDP and national income accounts are among the great inventions of the 20th century and are considered as an important indicator of development and economic growth in a country (BEA, 2012). However, there are still debates about the calculation of GDP, which is considered to be the best measure of the market value of goods and services produced. According to Samuelson, one of these debates is how to treat the economic product produced by housewives. In his textbook "Economics", first published in 1948 and continued to be republished until its 20th edition in 2019, Samuelson raises his paradox about "the level of GDP when a person marries her lover or her gardener" in the last part of the unit titled **"Measurement of Economic Activity"** (Samuelson, 1968 and 2009). In the third part of the study, this issue, which has still not been resolved despite the contributions of Scott as a second author alongside Samuelson in the later editions of the book and Nordhaus in the latest editions, is addressed.

In the literature, in accordance with the set of values (referred to as the paradigm) of the current capitalist economic system, labor is valued to the extent that it is employed and there are not enough academic studies that measure the contribution of non-employed laborers to GDP. Women's labor, which is largely excluded from the labor-employment relationship, is also deprived of the personal rights and wage income provided by employment. The inequality created by this deprivation is defined as "gender inequality" in the literature. In the history of economic thought, starting from the classical understanding to the Post-Keynesian understanding and including the Real Business Cycle Theory, which associates the basis of business cycle fluctuations with growth, the phenomenon of labor has been addressed with a focus on employment.

In many studies on these developments in economic theories, emphasis is placed on women's labor and the inequality experienced, but no analysis/suggestion is made beyond the concept of "employment" as a paradigm or outside the labor-employment relationship. However, from a heterodox economic approach, it is observed that a new paradigm is emerging that may be related to women's labor, in which the limited rationality of behavioral economics and heterogeneous individuals constitute today's human profile and concepts such as unpaid labor are sprouting. In economics, it is thought that "we need a heterodox approach that does not treat labor as a commodity and sees it as the greatest purpose not only of economic life but of all of life" and that labor should be removed from the status of commodity in discussions on labor and value (Bocutoğlu, 2012: 6).

For the concept of labor in economic theories, in line with this purpose, GDP accounts have been chosen as an example and a model has been tried to be created by using data from Turkey to include the production of housewives who use their labor as domestic workers. For these reasons, this study is important both theoretically and in terms of social utility: First, theoretically, it is important in that it envisages a new paradigm for evaluating labor independently of employment, which is the basic assumption of the study. Secondly, as a social benefit, it is important in terms of contributing to the debate on whether it is appropriate to include the labor and products of housewives in the economic system or whether they should continue to be ignored.

In the conceptual analysis, in the first and second sections of the study, the concept of labor is questioned in terms of its current meaning and the developments in economic growth calculations for measuring labor. In the third and final part of the study, the effect of housewives' labor on economic growth as an economic value will be measured in the case of Turkey through a model constructed using TURKSTAT data on GDP for the selected year and the Atkinson Inequality Scale.

In the first section, the concepts and theories related to labor power in growth indicators focus on whether labor is provided for paid or unpaid. The theoretical analysis in this section will examine the development of the concepts of labor and women's labor in the theoretical literature as accepted in orthodox and heterodox economics. In the second section, the calculation methods of GDP, which is accepted as a labor and growth indicator in economic growth models, will be analyzed. Taking into account the interrelationships between the concepts of economic growth, women's labor and GDP in growth theories; methods for calculating GDP, which is accepted as an indicator of economic growth, will be introduced. As a result of the analysis of the first two chapters, in which the existence of women's labor in economic theories, growth and GDP calculations is questioned, the labor-employment dependency and the unpaid labor supply of housewives against the labor demand of the family are tried to be explained. The concept of housewives' labor, which is defined as the independent variable of the study, is defined.

In the literature review, the third part of the study, the theories and policies to increase women's labor force participation in developed countries and Turkey are examined and it is shown that the first two sections of the study confirm the existence of labor-employment dependency in the current economic functioning. In the third section, the theoretical expectations of the study are presented and hypotheses are formulated. In order to support the theoretical analysis of the thesis, a sample calculation is used to show how the labor of housewives, who are considered as unpaid labor and constitute a significant portion of the population, affects GDP and social welfare. In this calculation, GDP, which is calculated by income methods, is chosen as a sample calculation tool to measure the contribution of the labor of limited rational individuals to the economy. In the calculation of GDP, a model was tested to include the labor of housewives. In the first stage of the model, in the case of Turkey, an income/utility gap is calculated for the current situation with the Atkinson Inequality Scale using the selected TURKSTAT employment data for 2014. In the second stage of the model, the income gap created by housewives is calculated and added to the 2014 GDP which is calculated using the income method to calculate the income/utility effect for that year. The results obtained in both stages are compared to show the economic value of housewives' labor.

In the last part of the study, a different method that can help GDP calculations was tried by taking into account the labor of housewives, which is unpaid labor; thus, the fact that labor is an economic value was tried to be de-gendered. The results of the application are combined with the theoretical analysis conducted in the first two sections of the study and presented as theoretical and empirical findings. The results are evaluated from the perspective of economic philosophy in the context of new approaches to the phenomenon of labor and women's labor, the place of labor in the economy, economic growth criteria and market criticism, taking into account the theoretical expectations of the study. Whether a woman is a hired laborer or an unpaid domestic worker, like any other labor owner in the labor market, when she stops supplying labor due to wear and tear or death, she is replaced by another labor force. Therefore, it is the responsibility of economics to ensure the visibility of the continuous demand for housewives' labor by households and the production that is generated by meeting this demand.

This thesis, which attempts to shed light on women's labor, which has been an economic asset since the beginning of humanity's progress towards equality between men and women, but has remained in the darkness of capitalism, is thought to contribute to economic welfare. Thus, with the confidence in science, it is expected that it will serve to eliminate the labor turmoil in social life and to ensure economic justice.

# CHAPTER ONE

# 1. THE CONCEPT OF LABOR AND WOMEN'S LABOR IN ECONOMIC THEORIES

Since the pre-classical economic theory, the concept of labor has been associated with the concept of value. According to the "just price" approach advocated by Aristotle and Plato, which was generally accepted in Antiquity (750 BC - 500 AD), an explanation was put forward on the basis of the concept of value within the framework of the prevailing understanding of justice of the age. Aristotle divided the concept of value into value of usebality "use value" and value of exchangeability "exchange value" and expressed value in terms of the ratio of the things they are exchanged for. Therefore, the economy, first in the Mercantilist period and then continuing until Adam Smith, is based on the understanding that the owner of labor and the user of labor are different people. This difference appears as a determining difference in rational decisions such as gender-based division of labor investment based production or production based on supply and demand.

In the economic literature, it is known that in each of the micro and macro theories[1], "labor supply" does not take into account female labor supplied by households for household chores. The capitalist economic system, both in terms of production theories and consumption theories, is based on the assumption of producing surplus value. Therefore, in economics, being a laborer is characterized as a person who supplies labor and receives wages in return. This approach can be explained as a situation that science ignores for labor suppliers whose wage condition is not met.

## 1.1. LABOR AND WOMEN'S LABOR IN ORTHODOX SCHOOLS OF ECONOMICS

As mainstream economics, Orthodox economics, in the historical period from the 1870s to the 1920s, criticized especially the labor theory of value of the Classical school. Orthodox economics, which made radical changes in economic thought, explaining many economic factors based on marginal laws and remaining loyal to liberal thought without breaking its ideological ties with the Classical school, has completed its historical period; but it is stated that its analytical process continues. Although orthodox economics found its scientific identity in the Neoclassical school, it underwent an evolution with the dominance of Keynes' ideas in the economy after the 1929 crisis (Kazgan, 2014: 118).

---

[1]Some of these topics are: Neoclassical work/leisure theory, preferences and indifference curves, budget constraint and equilibrium, utility maximization and work decision, non-wage income and working time: income effect, income and substitution effect (income/time graph), labor supply curves, derivation of the market labor supply curve, labor supply elasticity, factors affecting labor supply.

The theories developed in all schools of economics have been able to maintain their validity as long as they are in line with the interests of those in power, while gaining strength from self-validating processes and generally accepted practices. As it is known, economic policies shaped by the capitalist understanding take strength from the competition understanding of the market it creates; and in the light of this understanding, it is accepted that there are always two opposing sides: the defeated and the defender, the superior and the weak or the winner and the loser.

According to the capitalist understanding, everyone should strive to be on the winning side. The dominant paradigm formed by these policies used to explain all actors and all roles within the market understanding of the capitalist economic system is called orthodox economics (Eğilmez, 2013).

### 1.1.1. Mercantilist and Physiocratic Period

The Middle Ages and the Mercantilist period can be characterized as a period, far from being scientific about "value". However, it was reconsidered with scholastic and post-scholastic views; it is accepted that the value of goods can be measured through utility and scarcity; in other words, if the good is more useful, it has a higher value and an increase in the amount of supply will cause a decrease in the price and value of the good (Bozpınar, 2020: 74).

During the Mercantilist period, which prevailed from 1500 to 1800s, trade was carried out not with banknote money but with money minted from precious metals, gold or silver. Therefore, in this context where world wealth was considered fixed, being rich meant having more precious metals, and the enrichment of one could only be possible by the impoverishment of the other. The mercantilist economy is known as a trade economy that focuses on income and getting rich rather than labor. Merchants needed policies to ensure the entry of more precious metals into the country; this expectation was met by a strong, protectionist and interventionist state and the capital of the capitalists, who were merchants, largely consisted of stocks of goods to be sold. For them, the first way to make a profit was to exchange gold and silver in the United States for prices that changed over time due to the differentiation in the value of goods (inflation) as they moved geographically through trade. The second way to make a profit was the profitability created by the difference in the production conditions of the goods produced due to geographical differences (Kazgan, 2009: 47).

As can be seen, in the mercantilist period, the owner of labor; the person who produces and the person who gets paid for his substituent labor by putting what he produces on the market are the same person. Those who have economic power and direct the functioning are the merchants who stock the goods put on the market by the producer by buying in bulk according to the price and focus on making high profits to increase their capital by selling at different times or in different places. However, when the doors of the 17th century were opened, mercantilist commercial capitalism was struggling not to be replaced by industrial capitalism. Mercantilism can be said to have ended in the 18th century when industrial capitalism, which emerged as a result of technological inventions led by England in the 18th century, changing the forms of production and increasing industrialization with the effect of state interventions encouraging exports, replaced commercial capitalism.

Unlike the Mercantilists, the Physiocrats considered only land as the "creative" factor of production, while they considered all activities such as trade, barter and intermediary services that emerged after production - although useful - as "sterile" and did not recognize them as economic surplus value. In the Physiocratic period, when natural laws prevail, there should be no state intervention and the continuity of the natural order depends on the continuity of private property (Bilgili; 2015: 25).

The Physiocratic philosophy, which was caught in the turmoil that began with the transition from commercial capitalism (Mercantilist period) to industrial capitalism (Classical period), actually takes its source from nature laws. In this understanding, it is accepted that people are born with the same "human" nature and therefore should have the same rights and freedoms. The right to property and individual initiative are considered among the most natural rights that equalize people. There is a need for a social order suitable for the realization of these equal rights and freedoms that everyone naturally possesses, which can only be provided by the laws of nature. Capitalistic freedom is based on private property, individual initiative and the market is shaped by these structural principles; these principles are the laws of physics that shape the structure and functioning of societies, known as the regular and immutable laws of the material universe, as laid down by Newton. The idea that societies are also subject to the law of nature in their internal functioning and are governed by a "good and rational God" in natural harmony, led to the acceptance of the assumption (theoretically explained later by Adam Smith) that societies governed by an invisible hand for specific purposes could be in equilibrium. This acceptance required passive rationalism to identify and obey the natural laws of society, which in turn led to the conclusion that these were the laws of the market economy. The fact that natural harmony is seen as harmonizing individual and social interests has led to the discipline of the natural functioning of laws and the discipline of non-interference in this functioning, which would be useless even if it did occur. As a result, based on nature laws, the most appropriate economic order for the structure of society was accepted as "Laissez Faire", that is, a legal system that does not intervene in the market but only ensures the free functioning of market laws (Kazgan; 2009: 55-58).

It is stated that the concept of economic value was analized, on the one hand from the perspective of commercial capitalism, shaped according to the principle of "buy cheap, sell expensive", on the other hand as a subfield within moral philosophy with the scholastic method of analysis, has been going on for centuries and had its beginnings in the period of F. Hutcheson (1694-1746), Adam Smith's teacher (Bozpınar, 2020: 75).

As an element of production and trade; theoretical explanations about the concepts of labor value were put forward by the Classics.

### 1.1.2. Classical Economic Theory

The conception of trading goods for wealth in exchange for more precious metals, a holdover from the mercantilist period when trade was conducted in precious metals, led to the exchange of goods for money only; which in turn led to their gradual commoditization. Economists who wanted to determine how the prices of commodified products were formed in this capitalist formation used reference criteria to explain how the values of the goods themselves were formed. The first accepted explanations about goods and their economic value were published by William Petty in his book "A Study of Taxes and Contributions" published in 1662. According to Petty, the criterion that determines the prices of goods that change under the influence of "supply and demand" is the value of goods and the element that determines value is labor (Koç, 2014: 4).

According to the English philosopher John Locke, who lived between 1632-1704 as a pioneer of liberal thought, it is accepted that labor is the most fundamental component in the creation of an economic value (Bocutoğlu, 2012: 69). In his book "Second Essay on Government", Locke, who aims to explain the inequality of property, states that labor is necessary for the aquirement of property. The rules such as the prohibition of accumulation, the necessity of limiting political power and the validity of law in every field, which Locke put forward in his theory known as the social contract, have been accepted for centuries (Arslan, 2013: 195-197).

Another pioneering work on labor and value was done by Richard Cantillon (1680-1734), an Irish banker, in 1755 and published in his book "Essai sur la Nature du Commerce en General". Cantillon, who made a distinction between the values that must be possessed for the laborer to be able to produce and the output "produced by labor", drew attention to the fact that the minimum living level is affected by the element of "tradition" (Koç, 2014: 5).

Adam Smith (1723-1790), who was influenced by Aristotle and Roman philosophy, aimed to determine the value of goods and tried to explain the relationship between labor and value by arguing that the use value and commercial (exchange) value of goods are different from each other. According to Smith, the dependence between labor and value started in the primitive society period when the concepts of property and capital did not exist. Because there was no private property at that time, each individual had to meet his own needs. The concept of price, which emerged as a reference to the concept of value in the period of industrial capitalism when the concept of property and capital accumulation developed, is seen as different from 'value' according to Smith, due to its ability to change over time (Smith, 2008: 31-55).

Although the first scientific studies in the field of economics were conducted about half a century before Adam Smith's book "The Wealth of Nations" published in 1776, Smith is considered to be the founder of the science of economics and the explainer of the theories that labor is the determinant of value (Bozpınar, 2020).

In his book, Smith took a more systematic approach to the issues of labor and value and tried to reveal the relationship between the prices and values of goods and how the values of goods are determined in the process of the formation of prices of demanded products in the capitalist environment. According to Smith, "the true measure of the exchange value of all goods is labor" (Smith, 2008: 31-96).

According to Smith, in a capitalist economy, where capital accumulates in certain hands, production is carried out for the purpose of exchange (trade) and land can be owned. In such a society, there are three different interest groups. These are capitalists with capital, workers with labor and landowners. At the end of the process of production of goods and services, these interest groups have three basic types of income known as profit, wages and rent (Smith, 2008: 304-310). As can be seen, the economy is at full employment and there is no segment of society that does not work and has no income[2]. This is because, according to Smith, the part of production that is in excess of the need (not demanded in the market) belongs to the one who produces this surplus, that is, the owner of labor-power.

According to Adam Smith, labor is "the first price paid for all things; the real purchase money". Therefore, the vast majority of the world's wealth is purchased not through gold and silver, but through labor. Smith explains that "labor is the true measure of the exchange value of all commodities" based on the assumption that labor must be expended for any commodity to have value. Labor, as a means of measurement that allows the value of commodities to be compared with each other at all times and under all conditions, is both the universal and the only criterion of value formation (Smith, 2008: 70-118). Smith's scientific approach to the relationship between labor and value is known as the "labor theory of value".

William Thompson, on the other hand, in his book "An Inquiry into the Principles of the Distribution of Wealth" published in 1824, argued that all products created by labor are the right of labor. Based on the theory that labor is the sole creator of wealth, Thompson argues that the ownership of the product produced should belong to labor. Thompson, who attributed political consequences to the Labor Theory of Value, argued that profit and rent (rental income) consisted of stolen resources taken from labor without the permission of the labor owner (Koç, 2014: 47).

The reason for the existence of unanswered questions and debates in Adam Smith's labor theory of value, and thus the reason why the classical theory is still valid, can be sought in the fact that it has self-interested political implications. It is thought that during the progress of the capitalist system, Classical economists - probably for the sake of profit maximization - constructed theories to justify the liberal economic doctrine. Therefore, in order to understand the liberal economic doctrine at this point, it is necessary to first understand the theories put forward by the Classical economists, and the old books are reopened again.

---

[2]The capitalist system's assumption that "the economy is at full employment" is based on this understanding. This issue is discussed in the second part of the study under the title "Labor-Employment Dependency and the Invisible Labor of Housewives".

As a matter of fact, Adam Smith, in his first explanations on value in his book The Wealth of Nations (1776), states that "labor is the only criterion that determines the value of commodities" and that "the relative prices of commodities generally depend on the relative amounts of labor required for their production" and that there is a difference between the real prices of products and their nominal prices (Smith, 2008: 30-42). As it can be understood from this point, Smith's explanation that labor output is evaluated differently when it belongs to the owner and differently when it comes to the value of the commodities to be exchanged, shows that Smith's view that labor is the only measure that determines the exchange value in the pre-capitalist period cannot be applied in the capitalist economy (Kazgan, 2009: 50-75).

Smith explains the profit of the entrepreneur who accumulates goods or capital not with the theory that considers labor as value, but with a theory based on the addition of profit to labor (adding-up theory of value). Defining the savings remaining after the freight cost of the goods and workers' wages are paid as the entrepreneur's profit, Smith argues that profit, like value, is a measure of labor;"*It may perhaps be thought that the profits on the stock of goods are another name for the wages of a different kind of labor, namely, that of inspection and management. But they are quite different from wages, and are regulated on entirely different principles; there is no proportion between the amount and difficulty/skill of this inspection and management work (labor), which is supposed to exist, and this wage. They are always regulated by the value of the stock of goods used; they are larger or smaller according to the size of that stock...*" (Smith, 2008: 53).

Smith continues this explanation with an example of how the costs and profits of employing workers in two different industries, the number of workers employed and their productivity will determine the amount of profit. When this approach is examined more carefully in today's bright light and adapted to a family economy, we see the existence of women's labor, which Adam Smith did not see in his time and no one cared about. It is understood that the labor of a mother as a housewife and the labor she spends on supervision, management and even production for her dependent husband and children is also an economic value and should be compensated. Heterodox approaches to the subject are discussed in the second part of the study.

After Smith, the first theoretical explanations on the relationship between labor and population were made by Thomas R. Malthus in 1798 in his book "An Essay on Population". Malthus' population model is based on two assumptions: the first is that human beings need to be fed in order to survive; the second is the continuity of the dependence between the sexes. For the first time in economic theory, Malthus used an expression that drew attention to the relationship between the sexes; he predicted that, due to population growth threatening the labor input, the entire working class, which constitutes the labor factor (men and women), would be dragged into the risk of misery. Malthus, in the model he developed by considering the increase in population, argues that the subsistence level of market wages will not be able to eliminate unemployment even in the long run, as it will not be able to convince workers. Because in the long run, the surplus in labor supply due to the continuous increase in population, causes wages to fall to the subsistence wage level if the capital/labor ratio is not high enough (Bozpınar, 2020: 79-80).

One of the two forces that Malthus drew attention to, with the assumptions in his population model, is the power to reproduce (population growth) and the other is the power to produce food; and the power to reproduce is higher than the power to produce food. According to Malthus, the geometrically increasing population must be strictly controlled against the arithmetically increasing food production (Malthus, 1798: 6-34). So much so that even in England, as a country before industrialization, it was considered impossible for the rich to create wealth at their own standards by supporting their children (sons), which were too many due to the continuous population growth. Despite this impossibility, it is stated that the possible job opportunities were used for the sons who had power in the social hierarchy. So the sons of the master ended up becoming laborers, the sons of the merchant became artisans, and the sons of the large landowner became small landowners, while for daughters, the issue was limited to marrying as late as possible (staying at home) and producing children (Gövdere and Türkoğlu, 2016: 429).

As understood from Malthus's approach, the necessity of keeping women under control in order to keep reproductive power, i.e. fertility, under control in order to control population growth was attributed to economic reasons in the 1800s. Because in labor-intensive jobs, men are preferred to women as muscle power, and since it is the woman who gives birth, it is more possible to control women in the labor market. This point of view, which influenced the social culture of the time, despite the technological developments (machine-intensive production) that saved (labor-intensive) production from arithmetic increase and the development of birth control methods that controlled the geometric increase in population growth, is thought to continue influencing today's employment policies and causes women's labor force to be evaluated as cheap or unpaid labor.

According to Malthus, who feared the uprisings that might occur if the impoverished society could not be controlled in the face of increasing population: "If a man born into poverty cannot obtain her/his needs necessary to survive from his family and if society does not need him to work, he/she has no right to demand even the least of food; for he is a surplus..." (Malthus, 1798: 63).

The spreading of ideas of liberty and justice together with the French Revolution began to worry the capitalists in England, who made their living by keeping the people poor, and forced the rulers to improve the living standards of the people. When Malthus's book on population was published in 1798, the "William Pitt Act" (Bozpınar: 2020), which was put into practice for crowded and poor families, is therefore known as the Poverty Act; an example of how scientific explanations can serve political interests; wherewith it can be observed that Malthus pursued an approach that serves interest relations rather than an objective scientific approach.

As it is understood from Malthus's explanations, women's "near expulsion from society for a moral crime for which men are now almost unpunished is undoubtedly a violation of nature law based justice ". Women, along with the elderly and children, are defined as the most miserable segment of society and are seen as responsible for the population growth and poverty that starts within the family (Malthus, 1798: 10-13).

The woman, who is accused of weakening the nutritional power of the society with the population increase she causes, has to prefer to do housework instead of working in a second job due to the increase in the number of individuals in the family and has to work for subsistence, that is, for a full stomach. Alongside this understanding, which has continued until today, women are considered as a cost-creating factor, and the surplus value they produce while raising labor force (human resources such as workers, soldiers, etc.) for the country is totaly ignored, and they are tried to be "employed" in a second job outside the home with the necessity of "at least being useful". Women's economic freedom, as it is for every human being, is seen to be compensated for their labor.

Regarding the control of population, Malthus suggested the closure of institutions such as church aids and guilds, which ensure that the poor are constantly fed with aid, and the opening of free shelters by delaying marriages, stating that these measures lead to misery and immorality. Malthus, who thought that the improvement in living standards due to real wages and the decrease in mortality rates increased both the population and the labor supply, could not find a complete solution to the issue, although he thought that the population could be controlled with a reverse mechanism (Malthus, 1798: 65-80).

Marx made the most accusatory criticism of Malthus's population theory. Marx harshly criticized the naturalization of population growth as a result of the crises created by the capital owners, seeking to increase their profit, which is surplus value, in order to reduce costs for their own benefit. According to Marx, it is capitalism that creates poverty (Marx et al. 2008: 162-164). Since Marx's criticism is considered a new and different approach to economic theories, it is included in the section on heterodox schools of economics.

David Ricardo (1772-1823), another theorist in classical economics, accepted Malthus's analysis of population and wages and proposed that a stable wage level could be achieved through a stable population balance with the grain model he established by developing Adam Smith's labor theory of value. In 1817, David Ricardo concluded in his book "Principles of Political Economy and Taxation" that the exchange value of commodities (goods) is determined by the amount of labor they contain (Ricardo; 1817; 27-50). Unlike Smith, Ricardo makes separate explanations for his labor theory of value; both for the pre-industrial period and the post-industrial capitalist period.

In Ricardo's labor theory of value, which is based on agriculture, value is created as a result of labor or scarcity. Ricardo characterizes scarce goods that cannot be reproduced as "monopoly goods" and does not include them in the value analysis. With this approach, Ricardo's assumptions about the value created as a result of labor is that the economy is based on agricultural production, the capital/labor ratio does not change, labor can be calculated as homogeneous or uniform, and landowners do not contribute to production. However, with the transition to industrial capitalism, where capital accumulation and private ownership become more pronounced, Ricardo's labor theory of value acquires a new dimension; since the changing population and technological developments force agricultural production, which is the only production channel of the economy, to open unproductive lands to production. In this case, the price determined by labor, capital and rent as factors of production in the process on fertile lands, requires a pricing that includes only labor and capital for products on unproductive lands that require higher costs. At this point, which was not clarified by Smith, Ricardo, while determining the price depending on these two variables, proposes a common measure of value in which capital can be expressed in terms of labor. The theory that defines labor as direct labor and capital as indirect labor and that value emerges as the sum of direct and indirect labor is known as Ricardo's labor theory of value (Bocutoğlu 2012: 152-159).

However, Ricardo also failed to achieve his goal in determining an immutable criterion of value. Despite the increase in population, he made some suggestions to ensure the welfare of workers and to satisfy the owners of capital without state intervention. Accepting Malthus's proposal on population, he suggested population planning in order to raise the subsistence level of workers and thus ensure rapid capital accumulation (Ricardo, 1997: 36-45).

Another approach to labor that is evaluated within the Classical Economic theory is the Alternative Cost and Implication Theories. Jean-Baptiste Say (1767-1832) and William Nassau Senior (1790-1864), who did not find Ricardo's labor-value theory convient, created a division problem between labor and capital, which Ricardo described as labor - indirect labor, incompatible with the essence of the classical philosophy of the capitalist system, conducted studies suggesting that value does not originate from labor (Kazgan, 2009: 77-79).

In his argument, which he put forward as the Alternative Cost Theory, Say defines a value in which labor, capital and land are of equal importance, and the value of the good changing according to preference can be measured by the value of the non-preferred good. Ignoring the importance of the difference in the amount and duration of the labor factor spent for the preferred and non-preferred good, Say, with this approach, actually emphasizes capital and rent while devaluing labor. In his Theory of Distress, William Nassau Senior argues, as Smith did, that what determines value is the amount of effort and distress expended to obtain it. In other words, the distress of the capitalist in the process of saving and investing his income by squeezing himself to make a profit and the distress of the laborer in the production process are the same, equivalent. In this case, it is noteworthy that Say and Senior were biased in their approach in order to protect the interests of the capitalist, and not the laborer (Bocutoğlu, 2012: 160-165).

### 1.1.3. Neoclassical Economic Theory

William Stanley Jevons (1835-1882) admired Senior for opposing Ricardo's theory and went further; he criticized Ricardo for leading science away from the capitalist approach to labor and indirect labor and accused him of creating confusion in science. In his book "The Theory of Political Economy", Jevons argues that it is worth spending labor for a product that has a lot of benefit, but a value cannot be created just because labor is spent; there must be a benefit that emerges as a result of labor and this benefit must be measured correctly. According to Jevons, utility is not mathematically measurable, but it is an ordinal value and its ordinal value decreases as utility is obtained. Thanks to Jevons' diminishing marginal utility approach, there is a relationship between utility and its commercial (exchange) value. Jevons' diminishing marginal utility approach also solved the "water-diamond paradox" posed by Smith, who could not establish a relationship between utility and commercial (exchange) value. Jules Dupuit, a French engineer, and Gossen, a French engineer, made a similar approach to diminishing marginal utility. Jevons applied the concept of marginal utility to the general theory of rational choice. Changes in preferences that depend on price also affect utility. According to Jevons, an increase in the quantity of a product will decrease the marginal utility of that product, while a decrease in the quantity of an alternative product will increase the marginal utility of the alternative product and this situation will continue until the prices of both products are equalized (Bocutoğlu, 2012: 170-173).

Carl Menger (1840-1921) was another theorist who based the concept of utility as a criterion for value. In addition to the concept of marginal utility used by Jevons to explain exchange value, he explained it together with total utility, but did not use any mathematics. In his book "Foundations of Economics" published in 1871, Menger made subjective explanations about what value is and how to measure it. The value of a good can vary from person to person and the preferences of the individual are decisive (Menger, 2007: 145). In the price approach determined according to the behavior of the individual, Menger classified the goods that meet the consumers needs as low-grade goods and the goods that investors use in production as high-grade goods; therefore, he argued that marginal and total utility can only be applied to consumption goods. The price of a consumption good is determined by the total utility and marginal utility of that good, while the price of production goods is determined by the potential utility of the consumption good produced from that production good. Menger argues that the exchange value (profit or interest) of a production factor can be determined in the future by imputing its use value to the present value of its production factor (Menger, 2007: 114-165).

Thus, according to Menger, who rejected the labor theory of value, the value of labor is determined by the product, similar to Jevons' idea. It is noteworthy that Menger begins his presentation of marginal utility by defining what a good is and how different types of goods can be related to each other. These two discussions are considered important for Austrian marginalism as they foreground several central themes. Menger argues that for something to be considered as a good, it must meet four conditions:

1. There must be a human need for the thing.

2. The thing must be "capable of being brought into a causal connection with the fulfillment of this need".

3. Humans must know about this causal connection.

4. We must have enough authority over the thing so that we can use it to fulfill the need.

These conditions listed by Menger focus not on the object itself, but on its relationship with people and on human needs, knowledge and ability to dominate the thing in question. According to Menger, what makes something good is not a property of the good itself, but "only a relation between certain things and people". Attributing the property of being "good" to the relationship between something and people, reflects the distinctly subjective character of Austrian marginalism (Horwitz, 2003: 248).

In neoclassical economics, although the rules of classical economics such as perfect competition, full employment, Say's law and the effect of the invisible hand are accepted, it is argued that there may be deviations from perfect competition; it focuses on the principle of marginal utility and argues that the phenomenon that creates value is utility. The individual's choice not to work in the labor-value theory in classical economics is explained by the marginal behavior of the individual in the utility-value theory. The unemployment of an individual is his/her own choice; he/she can choose to work instead of being idle, until he/she reaches a certain level of utility (substitution effect) or she/he can choose to be idle instead of working, after reaching the desired level of income (income effect).

According to neoclassical economists, the determinants of price and wage are the same. In other words, determining the value of labor according to the supply and demand conditions of the market means that labor is commodified like a manufactured good (Bocutoğlu, 2012: 10). In the 19th century, expanding economic activities diversified the market phenomenon and the concepts of labor demand and labor supply to produce goods and services inevitably strengthened and increased the relationship between labor and employment. In fact, labor has become the most important component of the capitalist market as a commodity that is supplied by households in return for wages and demanded by firms to be paid for.

The development of the market phenomenon has also diversified the approaches to the amount and purpose of the inputs used in production, how they will be transformed into returns and how the output will be divided. J. S. Mill (1806-1873) argued that in contrast to the universal laws of production, distribution is universal.

J. S. Mill (1806-1873) argues that, in contrast to the universal laws of production, there are no universal laws of distribution and that distribution should be made by social consensus. J. B. Clarke (1847-1938) is known to have contributed to the issue of distribution with his "Theory of Distribution Based on Marginal Returns". It is stated that the way to be followed in calculating the amount that will fall to the share of other factors of production besides the labor owner, that is, how the distribution can be made, is open to discussion (Bocutoğlu, 2012: 67).

If the conception that emphasizes the issue of distribution as the fundamental problem of economic science is handled with the Neoclassical approach, including the necessity of protecting property rights, it is better understood, why the basic decision units, involving firstly, capital owners and secondly, working men with the ambition of acquiring property are directed to acquire property. Because acquiring more property means having more rights in the market. In such a class society, of course, women who work for their families cannot acquire property.

In the neoclassical period, based on Jevons' approach that "the value of labor does not determine the value of the product; the value of the product determines the value of labor", labor was transformed into a "commodity" whose value was determined in the market environment.

### 1.1.4. Keynesian Economic Theory

John Maynard Keynes explained his views on the solution to the 1929 Crisis in his book "The General Theory of Employment, Interest and Money" published in 1936. In his explanations known as the General Theory, Keynes argued that the imbalance between supply and demand that emerged in 1929, which he characterized as a lack of demand due to unemployment caused by the contraction in production, could be solved by increasing demand through increasing public expenditures; that is, by government intervention. Keynes argued that it would be possible to bring the markets into balance by intervening with the visible hand of the state instead of the known invisible hand of the Classical theory, which was insufficient to solve the deflation and unemployment problem caused by the Great Depression (Keynes, 2018).

While writing his book, Keynes admitted that the choice of units of quantity appropriate to the problems of the economic system as a whole, the role of expectation in economic analysis, and the definition of income were the three issues that most hindered his progress and therefore created confusion that he could not comfortably express until he found a solution to them (Keynes, 2018: 33).

While accepting the general theories of the classics, Keynes does not accept the explanation that the market will be in equilibrium at "full employment". He argues that men would become involuntarily unemployed if there is a small increase in the prices of goods and both the total supply of labor and the total demand for the product are greater than the current volume of employment. Thus, he provides a new definition of "involuntary" or "voluntary" unemployment, different from the "voluntary" unemployment in the classical theory (Keynes, 2018: 14-15). That is, in a "market" whose conditions are determined not by the owner of labor but by the owners of capital, there is a possibility that a worker who wants to work at the current wage level cannot find a job. This unemployment caused by market conditions is defined by Keynes as "involuntary unemployment". As can be seen in these explanations, Keynes, like the Classics, considers the labor phenomenon as dependent on the condition of employment.

In the introduction to the 2018 edition of Keynes's General Theory, Paul Krugman emphasizes that Keynes was not a socialist and that he came to the rescue of capitalism, and that Keynes himself found his theory "moderately conservative" in some respects, owing to the fact that the book was written at a time of mass unemployment, waste and suffering on an incredible scale. According to Krugman, Keynes's attempt to rescue the failure of capitalism through state intervention in the economy is the restoration of economic reason as the nationalization of the means of production. It is noted that many British and American intellectuals, who had no particular antipathy towards markets and private property, became socialists during the Depression years because they could see no other way to correct the colossal failures of capitalism (Keynes, 2018: 27).

### 1.1.5. Neoclassical Synthesizing Keynesian Economic Theory

The Keynesian approach such as perfect competition, full employment, Say's law, the invisible hand and marginal utility adopted in neoclassical economics, that is, the demand-side form that accepts state intervention in the economy, is called the Neoclassical Synthesis Keynesian economic theory (Bocutoğlu, 2012: 271).

This theory can also be considered as an indicator of the development of capitalism. This is because it is seen that the developments in the phenomenon of "money" and monetary diversity have moved to a stage where they should be evaluated in a separate market. In the Neoclassical Keynesian theory, while analyzing the relationship between goods and money markets with the IS-LM Model, the underemployment condition is also considered.

In the Neoclassical Synthesist Keynesian theory, while analyzing the relationship between goods and money markets with the IS-LM Model, it is seen that although the underemployment condition is also considered, it does not make a difference in the understanding of the evaluation of labor in a certain market and the condition of employment.

### 1.1.6. Monetarist Economic Theory

The theory, published by Milton Friedman (1912-2006) in 1976 in his book "Studies in the Quantity Theory of Money", is based on a purely monetarized economic functioning in which financial stability can be achieved by controlling employment and inflation.

Throughout the 1960s, Keynesians and orthodox economists in general, based on the Phillips[3] curve, believed that the government faced a stable long-term trade-off between unemployment and inflation. According to this view, by increasing demand for goods and services, the government could permanently reduce unemployment by accepting a higher rate of inflation. However, in the late 1960s Friedman (and Edmund Phelps of Columbia University) challenged this view, arguing that once people adjust to a higher rate of inflation, unemployment will rise again. Developing the use of the Phillips curve, he argued that to keep unemployment permanently low, it would require not only a higher, but also a permanently accelerating rate of inflation. According to Friedman, who argued that natural unemployment is a normal state of affairs in the economy, actual unemployment below the natural rate of unemployment leads to inflation, while above it leads to deflation (Henderson, 2018: 3).

[3] Although it had predecessors, A. W. H. Phillips' Phillips curve, which represents the opposite relationship in the United Kingdom from 1861 to 1957, showing that when unemployment is high, wages rise slowly; when unemployment is low, wages rise rapidly, is considered a milestone in the development of macroeconomics.

Taking a more moderate approach to the female population than Maltus, Friedman states that the reason for the increase in the natural rate of unemployment in the US society he analyzed is the "unemployment" of women between the ages of 13 and 20, who constitute a large part of the population and who work part-time. In addition to M. Friedman, Monetarism developed as a school of macroeconomic thought that emphasized long-run and short-run monetary neutrality, the distinction between real and nominal interest rates, and the role of monetary aggregates in policy, with writings by A. Schwartz, K. Brunner and A. Meltzer, and interest from outside the US, including D. Laidler, M. Parkin and A. Walters. It is noted that many free market advocates do not prefer to identify themselves as Monetarists (McCallum, 2018: 22).

It is noteworthy that monetarists focused on the formulas for monetary growth without labor or the value of labor in a monetary macro development environment. While treating labor as a market component, they explained the relationship between unemployment and the market and defined the unemployment of a part of the population in the market as the "natural rate of unemployment". They argue that the willingness of a person to hire his or her labor for a wage, while establishing a real wage level in the market, is also a determinant of (adaptive) expectations[4] about the future and the market.

In this case, it can be said that the Monetarist theory goes one step further than the commodification of labor by giving importance to wage income in the individual's decision to work and in making money a motivational tool and more important than labor. It is seen in the ongoing efforts to increase women's employment all over the world, that policies to reduce the natural rate of unemployment, which started in accordance with the monetarist theory, education, incentives and practices, shortening the time to find a job in order to increase the working population, are effective in directing housewives to employment.

### 1.1.7. New Classical Economic Theory

In the early 1970s, the monetarist opposition to Keynesian thinking began to be challenged by the New Classical macroeconomics of R. Lucas, T. Sargent, N. Wallace, R. Barro and other leading economists who argued that market economies spontaneously equilibrate and that public policies are ineffective in systematically stabilizing the economy. In this approach, it is argued that individuals and businesses, as owners of capital, act "rationally" in determining their expectations about the future according to the lessons of the past. It is stated that the New Classics developed the concept of rational expectations introduced by John F. Muth by using it in macroeconomic analysis (Yıldırım and Özer, 2014: 8).

---

[4]Adjustment of estimated inflation to current inflation.

In the New Classical economic theory, since prices and wages are always flexible, unemployment in the market is also voluntary. In other words, people are unemployed due to low real wages and unless the market is intervened, a price-wage equilibrium is formed that will eliminate unemployment. There can be no intervention in the market (Üstünel, 1990: 273).

In the New Classical Economic Theory, the phenomenon of labor is valuable under the condition of employment; since it is accepted that wages and prices are flexible; there is no involuntary unemployment in the market.

### 1.1.8.   New Keynesian Economic Theory

The theoreticians of the New Keynesian Economy are S. Fisher, E. Phelps, M. Parkin, J Taylor, J. Stiglitz, A. Blinder, G. Mankiw, G Akerlof, L. Ball and D. Romer, with M. Parkin as the namesake. By analyzing the micro causes of macro imbalances and taking into account aggregate supply as well as aggregate demand, they advocated rational expectations as in the New Classical theory. However, in New Keynesian economics, state intervention is necessary for markets to self-clean themselves. In the market, labor is employed for wages and the rigidity of wages is characterized as an imbalance. The state should intervene in this imbalance of labor markets (Bilgili, 2015: 309-311).

New Keynesians explained the issue of wage labor with the effective wage theory; the hysteria effect, insiders-outsiders and implicit contract theories, taking into account the effects of awareness and unionization in the market. In the New Keynesian Economy, labor is included in production on the condition that it is employed as a wage, and it is seen that the labor-employment dependency continues.

### 1.2. LABOR and WOMEN'S LABOR IN HETERODOX SCHOOLS OF ECONOMICS

By adopting pluralistic ideas, heterodox economics deals with different theoretical studies without adhering to only one approach (Tomanbay, 2019: 42). In heterodox economics, which has emerged with new developments by departing from mainstream economics, a social structure shaped by the past learned from rational-individualist and equilibrium approaches is at the forefront (Eren and Uysal, 2017: 137).

According to Lovie, there are many schools of thought within heterodox economics, most of whose members are openly opposed to neoclassical economics. These include Marxists, neo-Ricardians, neo-Structuralists (on development issues), Institutionalists, the French School of Regulation, Humanist or Social Economists, Behaviorists, Schumpeterian (also called Evolutionists), Feminists, Post-Keynesians and more. According to Lovie, heterodox economics is subject to two opposing forces. First, although heterodox schools are composed of competing schools, they are complementary because they focus on a particular aspect of economics. Second, there is also a tendency towards unity among heterodox schools, perhaps as a result of their small number, and many heterodox thinkers seek interaction and unity between their approaches. This is particularly true for American Post-Keynesians and neo-radicals (Marxists) working in macroeconomics and monetary theory (Lovie, 2006; 1-2).

According to Erkan, the reliability of the equilibrium models of orthodox approaches is questionable, given their failure to predict the crises of the global economy. In this case, heterodox schools with new approaches that are needed are listed as Historical School, Marxist School, Austrian School of Economics, Institutional Economics, Post-Keynesian Economics, Public Choice School, Feminist Economics, Neuroeconomics, Green Economics, Experimental Economics, Behavioral Economics, Evolutionary Economics, Evolutionary Game Theory (Erkan, 2016: 31-32).

At the end of the 1960s, in opposition to the orthodox theory, which was characterized as one-sided as monism, heterodox schools, which represent the pluralist or diversity-oriented approach, adopted theories, intertwined ideas and analytical studies in which cooperation rather than competition was observed. In "Directions for the Future of Heterodox Economics", editors Garnett and Harvey note that 21st century economists bring together topics ranging from mathematical to philosophical ideas, from critical to positive thinking and from pro-market concerns to socialist thoughts. In doing so, heterodox economists have much to contribute to economic theory through innovative connections, new lines of research and new dialogues (Garnett and Harvey, 2008: 2).

Unlike neoclassical economic theory, heterodox schools of economics, which aim not to treat labor as a commodity and to show that it is a very important value for the whole of life, are divided into traditional and new schools; but in this study, perspectives on labor will be analized from the perspective of historical developments, as is shown below.

### 1.2.1. Marxist School

Karl Marx (1818-1883), in his labor theory of value in Volume I of his famous work Capital, in which he explains his views against the capitalist system, defines labor power as the worker's latent capacity or ability to work and defines labor as the act of producing real value. In this distinction, labor is the labor that the worker uses for as much of the work as necessary to produce surplus value by renting it out. Labor forces, on the other hand, are in the vital continuum to be used at other times, and the value of a commodity is determined by the labor-time hidden in that commodity, which is socially valuable. Marx assumed that the organic structure of capital is the same between firms in each sector and across sectors.

With this approach, similar to the approaches of Smith and Ricardo, who remained most faithful to the labor theory of value, Marx finalizes the value of the product of a production as "labor". However, in Marx's labor theory, labor determines the absolute value of goods and services, unlike Ricardo's labor, which is the measure of the relative value of goods and services. The fact that the owner of the labor, hired by the capitalist who pays the price for his/her labor, has the capacity to produce more than the price paid to her or him; this is thought to be the reason why labor has been the subject of exploitation up to this day.

While explaining these thoughts on labor exploitation, Marx defined the concepts of labor-time, residual value and residual value rate, profit rate, transformation problem, capital accumulation and decrease in profit rates, centralization of capital and concentration of wealth in certain hands and class conflict; and made determinations on these issues. According to Marx, when the labor-time/labor-power ratio is greater than one, "surplus value" is formed and the time spent by the worker for the price he will receive in return for his labor power increases. In other words, Marx defines this situation of working more than one hour for the wage that would be received for one hour, as exploitation. With a similar approach, he proves the theory of exploitation through the ratio of unpaid labor to paid labor by proportioning the "fixed capital" used by the capitalist for investment to the "variable capital" that he pays as workers' wages (Bocutoğlu, 2012: 136-140).

In Marx's labor theory of value, the problem of transformation is the issue that he himself contradicts. In Volume III of Capital, published in 1894, it is stated that with the transition to industrial capitalism, as the capitalist's profit began to replace the residual value in calculations, the market price of goods began to be traded instead of their market value. In this case, Marx explains the problem that arises as the transformation of values into prices as the problem of transformation. The possibility that goods sold at their own value cannot be sold at their own value in differentiated industries due to the differentiation of the labor-capital composition, and that it is possible for capital to add value to the product, constitutes a contradiction that disrupts Marx's clarity that "it is only labor that constitutes the value of production". The fact that the prices of goods produced with the same amount of labor are varying, invalidates the "labor theory of value" in the pricing of goods (Savaş, 1974: 4-7).

The transformation problem is important in terms of the commodification of labor and the debate on its solution continues. Although the solution of this problem is a separate research topic, its importance for this study is to observe the use of labor as the unit of measurement used in the measurement of value.

In his book published in 1968, Sweezy defends the validity of Marx's theory by arguing that there will be no significant difference between market value and market price, which replaces market value. This is because, according to Sweezy, the reason for the problem is the use of labor as the unit of measurement. If labor is not used in the calculation of profitability, such a difference and thus the problem of value-price transformation would not occur (Savaş, 1974: 19).

Marx's determinations on labor by opposing the capitalist system seem to be against the capitalists and in favor of the owners of labor. However, Marx defines the female factory worker as "an individual who earns her bread" (Marx et al., 2008: 90) and defends women's rights by freeing them from domestic slavery and including them in the "working class" in employment. Instead of making a suggestion that women's labor, being the wife who produces at home, should be compensated by the social state, she is referred to as the family member to be nurtured by the person who is the labourer (the man) of the family. Marx can be characterized as a capitalist, because he devalued domestic work by seeing women's labor as a slavery under the auspices of her husband - the boss of the household - and included the cost of subsistence in the sphere of the man's livelihood, in other words, he evaluated labor as focused on the decisions of the owner of capital and dependent only on the condition of employment.

### 1.2.2. French School of Regulation

Until the first half of the 19th century, the world's population[5] was below one billion, but rapidly increased to over two billion in the early 20th century and is now approaching eight billion. In other words, the fact that the accumulations acquired to meet the needs of the population, whose growth has continued to accelerate in the last two hundred years, have begun to be used in applications where technology is used; which in turn gave birth to the necessity of regulations regulations in the field of economy.

The introduction of technology, known as Fordism, as a mass production method; the emergence of capital accumulation created by excessive profitability as a new regime of accumulation; and the need to meet the needs of the growing population.

While this development, which was initially pleasing for capitalists, actually depended on the realization of mass consumption, the crisis of 1929 showed that consumption, which depends on mass demand, did not materialize immediately. This crisis led to the highest unemployment rate in history and to the questioning of the labor-wage relationship, which had never satisfied the owners of labor. As labor became a mass producing and consuming force, it is stated that labor was organized through trade unions that sought improvements in working conditions, social rights and wages for workers (Harvey, 1989: 121).

It is stated that the theory of regulation consists of two basic concepts; one is the regime of accumulation and the other is the mode of regulation itself, and that it has an institutional functioning formed by the process–to ensure harmony between both. Forms of institutional functioning are defined as areas in which various reforms are carried out in order to reconcile both the stability of the newly formed accumulation regime and the regulations regarding the distribution of tasks among institutions with the accumulation regime, shaped by economic conditions in history. The harmonizing reforms between the accumulation regime and regulations are realized within five institutional forms (Hein et al., 2014: 34-36). These are;

---

[5] https://ourworldindata.org/world-population-growth
https://populationeducation.org/curriculum-and-resources/population-information/

Monetary regime regulating banking and credit relations,

The relationship between wages and labor, which regulates the position of labor in society,

The capitalist mode of competition, which can be monopolistic or liberal at times,

Whether the role of the state within the capitalist structure of the economy is regulatory and interventionist, and finally,

The country's position within the international regime.

While the reproduction of forms related to institutional functioning is made possible by the mode of regulation, the harmony of demand, production and income distribution expresses the regime of accumulation.

This effort of the theorists of the French School of Regulation is considered to be an important ~~as an~~ approach that tries to analyze the production and distribution paradoxes of capitalism in order to provide conditions of stability by creating institutions that develop rules and norms that provide a kind of protection against the instability of capitalism, which tends to create crises and obtain a profit from these crises.

### 1.2.3. Austrian School of Economics: 1950-2000

The Austrian School of Economics was founded by Carl Menger. In the school, which is also included in neoclassical economic theory, the doctrines formed by the Austrian marginalists have changed and the relative position of the school in mainstream economic thought has moved from the center to the extremes several times during its 130-year history. As it will be remembered, Carl Menger, in his 1871 Foundations of Economics, substituted subjective marginal utility for the Classics' objective cost of production as the theory of value. Later, F. Wieser's introduction of the idea of opportunity cost, emphasizing its subjective and ubiquitous character, and the work of E. Böhm-Bawerk, and the application of Menger's theory of value to the theories of capital and interest, seem to have developed on a path away from the mainstream. The precursors of the new heterodox generation, referred to as the Austrian School, are considered to be L. Mises and H. Mayer, who emphasized epistemic[6], ontological[7] and other philosophical themes[8] (Boettke and Leeson, 2003: 445).

However, in the period immediately after the First World War, the basic insights of Mises and Hayek were much less appreciated by other economists, and the main principles of the Austrian school, which the members of the fourth generation considered to be fully integrated into the mainstream, were explained by Fritz Machlup. Machlup emphasizes that Austrian economists were never uniform in their belief structure, arguing intensely among themselves over the relative importance of concepts and principles. Nevertheless, Machlup offers six principles, that is accepted by the economists trained in the Austrian approach (Boettke and Leeson, 2003: 447):

1. Methodological individualism. All economic phenomena can be traced up to the actions of individuals; which means that individual actions should serve as the basic building blocks of economic theory.

2. Methodological subjectivism. Economics takes human ultimate goals and value judgments as given. Questions of value, expectation, intention and knowledge are created in the minds of individuals and should be evaluated within this framework.

3. Marginalism. All economic decisions are made on the margin. All choices are choices about the last unit added or subtracted from a given stock.

4. Tastes and preferences. Individuals' demands for goods and services are the result of their subjective assessment of the ability of such goods and services to satisfy their wants.

5. Opportunity costs. All activities have a cost. This cost is the highest-value alternative foregone because the means for its satisfaction have been allocated to another (higher-value) use.

6. Time structure of consumption and production. All decisions are made in time. Decisions on how to allocate resources to consumption and production purposes over time are determined by individuals' time preferences.

Machlup presents the other two principles of the Austrian school, which he considers "highly controversial", as follows:

---

[6].Epistemology is the study of what distinguishes reasoned belief from opinion. It is the theory of knowledge, concerned with method, validity and scope.
[7]Ontology is the branch of metaphysics that deals with the nature of being.
[8]The source cites F. A. Hayek, Gottfried Haberler, Oskar Morgenstern, Fritz Machlup and Paul Rosenstein-Rodan as the fourth generation of Austrian economists, most of whom would make their academic mark in the US after the Second World War..

7. Consumer sovereignty. The consumer is king in the market. Consumers' demands dictate the shape of the market and determine how resources are used. Market intervention stifles this process.

8. Political individualism. There can be no political freedom without economic freedom.

These differences can be seen most clearly by looking at how an Austrian educated in the 1950s defined his school of thought and by comparing it with Machlup's understanding. Machlup's list of six principles is said to have been advanced by the theoretical contributions made by Mises and Hayek in the 1940s and to be the hallmarks of the Austrian approach to microeconomic theory. The Austrian position on macroeconomic theory can be summarized as follows: The Austrian School is in a different position vis-à-vis other economists because of the ability of contemporary Austrian economists to combine heterodoxy and orthodoxy. The school offers only microeconomic explanations and solutions to macroeconomic problems such as unemployment, inflation and business cycles. In the heterodox Austrian school it is important that the economic analysis of aggregate relations depends on individual choices. It is argued that the problem of aggregation in the relationship between the economy and its aggregate variables - although identified with Keynes' economics - is that aggregation masks the structural composition of an economy, and that the economist who seeks to understand overall economic performance must carefully examine the individual (Boettke and Leeson, 2003: 452).

### 1.2.4. Feminist Economics

When we look at the history of civilization in which the white man has been the ruler, it seems possible to understand why the person who is the decision-maker, the one with authority and property, literally the "master" alone as it is known in all cultures, is definitely a man and never a woman[9].

If housewives do not produce work, it can be thought that a capitalist patriarchal economic market has been transformed into a capitalist patriarchal economic market fiction where, on the one hand men, who produce and have other men producing for them, see in themselves the right to consume and manage everything, and on the other hand housewives, who do not produce but only consume as the target customer of the market. In order to eliminate the gender inequality that emerges when the rules of the developing civilization are shaped by the principle of equality, the reactions to the increasing extent of capitalism's exploitation and neglect of labor are called feminist economics within the heterodox economic understanding, which argues that women's labor and women's perspective should be taken into account in efforts to ensure economic prosperity.

---

[9]See Zubritski et al, 2007 and Engels, 2012

The individual, by socializing and organizing to meet his/her needs, which are considered to be unlimited, and by forming the state sustained by civil society, makes it possible for capitalism to develop and to attain an organic structure. It is stated that in the capitalist system, despite democracy, the woman remains a "domestic slave" confined to the bedroom, nursery and kitchen (Marx, et al., 2008, p. 207).

Despite its internal contradictions that can create crises of their own, capitalism teaches those whose rights it usurps, especially women, that they lack the ability to make appropriate decisions and that they should be grateful that their lives depend on the decisions of those who are smarter than them. In this way, capitalism rationalizes exploitation and convinces these people that it is an attractive and inevitable system (Hahnel, 2005: 77-78).

In the capitalist system, women who marry and devote themselves to housework, producing and raising children are allowed to donate their labor to the patriarchal system. For this reason, the capitalist system owes its accumulation to the labor of the housewife, who is uneducated and 'unskilled' with such general capacities, including friendship, docility, always being at the disposal of others, being sexually available, keeping everything in order, being frugal and modest, keeping herself in the background, having the patience and discipline of a soldier, enduring and being helpful in all matters and having "zero production cost" (Mies et al. 2008: 274).

The feminist understanding of economics argues that the economy, like many scientific, artistic and social structures concerning human beings, is shaped in a masculine way and characterizes the whole economy, which we know as the mainstream economy in the patriarchal system; the so called "male-stream economy". In the modeling and calculations of almost all economic approaches, especially orthodox schools of economics, the subject that deserves the term "male-stream economy" giving the priority to masculinity, contradicts the economic human profile known as 'homo economicus'. In written literature after 1990's, it is stated that with the developments in feminist economics, economics will become a discipline that reflects experiences of all people as it develops in a way that will make it possible for economics to be free from hierarchy and bias to include the economic efforts of all people, not just men (Işık, 2020: 130-134).

As Harvey points out, women are constantly made a part of the idea of "otherness" produced in capitalist society and made potentially unknowable. Thus, gender is added to the class exploitation of blacks, colonized peoples and minorities of all kinds, which causes various negative consequences for the society. It is stated that capitalism does not invent the "other", but it certainly uses and promotes it in highly structured ways (Harvey, 1989: 104).

As is well known, the transformation of labour into wage labour and its "thingification" as a factor of production gives the capitalist who buys this power with money certain rights over labour - regardless of what the labourer might think, need or feel. Feminist economics takes the approach that women should not be separated from men in the exercise of these rights; it advocates equal pay for equal work, increasing women's employment and equal rights with men.

### 1.2.5.  Post-Keynesian Economics

The term 'post Keynesian' was first coined by Eicher and Kregel in 1975 and subsequently by the Journal of Post Keynesian Economics in 1978 to refer to a new school of thought. Inspired by Keynes, post-Keynesians see capitalism as a system that encourages initiative and innovation. N. Kaldor, M. Kalecki and P. Sraffa are among these economists. The views of the post-Keynesians, which are also in line with those of the French School of Regulation, are closely linked to the work of the Institutionalists, inspired in particular by the ideas developed by T. Veblen and J. K. Galbraith. Thus, post-Keynesians believe that capitalism can be an efficient economic system, especially when it comes to income distribution and the provision of public services and public infrastructures to all layers of society, provided that deficiencies and excesses can be corrected by the state and supported by democratic institutions (Lovie, 2006; 131).

The theoretical basis of post-Keynesian economics is known as the principle of effective demand, which states that demand is important not only in the short run but also in the long run; and therefore the competitive nature of the market economy does not have a natural or automatic tendency towards full employment. Unlike the New Keynesians who adhere to the traditions of neoclassical economics, post-Keynesians accept that rigid or sticky prices or wages provide the theoretical basis for achieving full employment in the market (Arestis et al. 1999: 112).

Post-Keynesian economic functioning emphasizes "effective demand" as a concept closely related to labor and accepts Keynes' view in his General Theory that underemployment equilibrium can be achieved in the market due to involuntary unemployment. Keynes argues that even with perfectly flexible monetary wages and prices, there is no automatic mechanism by which the parties (employer-employee-government) can restore the full employment level of effective demand by making concessions to each other. In other words, following the logic of Keynes' general theory, it follows that in a perfectly competitive economy with freely flexible wages and prices, anything less than full employment equilibrium can exist (Davidson, 2002: 8). At this point where post-Keynesian economists meet Keynes, they adopt a more progressive view than Keynes, with more emphasis on pro-worker policies and redistribution (Harcourt, 2006: 6).

As a continuation of the questioning that began with Keynes' revolutionary explanations about the elements of economic equilibrium, Post-Keynesian economics shifts the issue of full employment in favor of working consumers, who constitute a large part of the population, and moves it to the demand for money, and accordingly to production and income distribution, and even to growth and development. As the users and carriers of money, it is seen that the owners of labor in employment play a more active role than ever before. Thus, these carriers also play an active role in the globalization of the capitalist economic system.

### 1.2.6. Institutional Economics

As an alternative to the market-oriented Classical and Neoclassical economic theories, it is stated that institutional economics, which is based on human-oriented approaches, has developed as an alternative to the market-oriented Classical and Neoclassical economic theories –owing to the fact that the effects of the psychological aspects of human beings began to be included in economic analysis (Doğruyol and Aydınlar, 2015: 266-267).

The reflections of the individual's preferences on the economy were first addressed by Adam Smith in 1759 in his "Theory of Moral Sentiments", before the publication of his book on the labor theory of value. In this work, Smith listed the forces affecting human actions and behaviors as self-concern, freedom, conformity to rules, industriousness and openness to innovations, and argued that the works they undertake for their own benefit actually affect the development of social welfare. However, it is stated that Smith's work gained meaning and was taken into consideration in the late 19th century when the focus was on human beings and their reasons for working (Kumcu, 2009: 276).

Similar to Adam Smith's approach, American economist T. Veblen (1857-1929), who investigated people's need to work and their psychological reasons, argued that people's desire to be liked and their curiosity to show off were at the basis of their economic decisions (Veblen, 1898 and 1918). It is believed that Veblen's sociological evaluations provide a theoretical basis for heterodox approaches.

According to Veblen, who is considered to be the founder of the institutional economics school, the institution consists of habits as well as of thought and behavior of the society. In his 1898 work "Why Economics Is Not an Evolutionary Science", Veblen describes neoclassical economics as incomplete because it is an economic functioning that does not take into account the thoughts and behaviors of society. Veblen finds the 'homo economicus' defined in the "ceteris paribus" conditions from which mainstream economics takes its data unrealistic and argues that economies full of rational people cannot change and develop (Veblen, 1898: 394)

In his book "The Theory of the Leisure Class", published in 1918, Veblen describes a leisure class, extensively composed of nobles and priests, in societies of barbarian culture, where industry was not yet present. In the early stages of economic development, unlimited consumption of good goods - ideally all consumption in excess of the subsistence minimum - is normally seen as belonging to the leisure class. Veblen criticizes capitalism, which he sees as the cause of the emergence of this class of idlers identified by Marx as holding surplus value and pursuing "conspicuous consumption". According to Veblen, this injustice in the social structure influenced the economic life of later times, shaped many traditions and even became a governing force. However, on the issue of economic injustice, he thinks that capitalism tends to disappear formally after the later peaceful[10] stage is reached with private ownership of goods and an industrial system based on wage labor or a small household economy; it will be abolished sooner or later in the course of development (Veblen, 1918: 2-38).

---

[10]Veblen uses the terms peaceful and semi-peaceful in response to a latent conflict created by injustice arising from class differences.

By taking the authority to use labor away from the capitalist and returning it to the laborer, the institutional economics approach promises to break the silence of women's labor that causes economic injustice and to give hope that economic justice can be achieved.

### 1.2.7. Behavioral Economics

The concept of bounded rationality is prominent in Behavioral Economics studies, which are considered to be pioneered by Herbert Alexander Simon (Simon, 1997). He argues that the rationality of the individual adopted in orthodox economics is unrealistic and that a more realistic definition is that real individuals are limited and cannot always be rational. Simon has devoted his life to explaining the nature of the thought processes that people use to make decisions; and in his book "Managerial Behavior" in 1997, he brings together his work on the issues that attracted his attention in economics in the 1980s and 1990s. He explains how to formally represent causal ordering in dynamic systems, the effects of new electronic technology on society, the effects of information systems on the economy, and economic issues. Presenting alternative models based on concepts such as "satisfaction of needs" as the acceptance of valid but not inexplicable and suboptimal choices and "bounded rationality" where rational calculation can guide human behavior to a limited extent, Simon demonstrates the need for more empirical research based on experiments and direct observation rather than statistical analysis of economic aggregates.

Simon first introduced the term "bounded rationality" in a 1957 briefing against neoclassical economics as a call to replace the assumptions of perfect rationality of 'homo economicus' with a concept of rationality adapted to cognitively limited effects (Wheeler, 2020, 1).

In his studies, Simon defines the bounded rational individual as an individual who can adapt the data he receives from his past experiences and environment to his own life. Considering that the rational individual of classical economics is not influenced by other individuals in determining his/her preferences, it is accepted that the bounded rational individual is more realistic. The fact that an individual selected from a community that is assumed to consist of homogeneous individuals in order to understand how he/she will behave under certain conditions behaves in a way that maximizes his/her utility actually confirms the rationality of this individual.

According to Simon, the individual who influences institutional structures in social understanding is also affected by them (Simon, 1997: 75).

In behavioral economics, individuals are defined as heterogeneous individuals who are open to learning and adaptable. An example of heterogeneous individuals having to learn and adapt to changing conditions in order to survive is women's silence as a solution to the labor exploitation that has continued for so long in the patriarchal economic system.

## CHAPTERTWO
## 2. LABOR AND HOUSEWIVES' LABOR IN ECONOMIC GROWTH CALCULATIONS

The "path followed by economic theory, the problems of production and distribution and the testing of their answers, the way of questioning and the interpretation of the results are based on the paradigm created by the patriarchal system. It is thought that this paradigm problem is at the heart of the current debates questioning growth, development and the welfare of humanity, which began with the shaping of human values. Considering that development is shaped according to the division of labor and that the division of labor is differentiated according to gender, it is stated that the labor of housewives, which is ignored in the capitalist system, should be redefined as a reason for the gap in development policies" (Kumcu, 2019: 2).

From Smith, who put forward the labor theory of value and said that the inevitable prerequisite for the value of goods must be human labor, to Ricardo, who emphasized the natural nature of the relationship between the labor involved in goods and the exchange value, and finally to Marx, who brought the labor theory to a logical line by making statements about women, the main reason for questioning labor is thought to be distribution. In order to make the distribution, it is important to calculate the whole.

### 2.1. ECONOMIC GROWTH and GROWTH MODELS

There are different views and models for determining economic growth. Analyzing the theories of the Neoclassical and post-Keynesian period from Adam Smith to Ricardo, Marx and Harrod, Hartcout states that by discussing endogenous growth theory as the 'new' growth theory and the Harrod model as the 'old' growth theory, one is back to square one or in fact, has never left Adam Smith's insights (Harcourt, 2006: 84).

Since the 18th century, it is stated that the concept of labor, whose exchange value was taken into account in the economy and whose use value was accepted as a prerequisite, was organized by becoming a mass producing and consuming force on the macro scale of the 20th century and turned into a means of distribution of growth (Kumcu, 2009: 275).

### 2.1.1. Economic Growth

Economic growth is defined as increasing the limits of production possibilities or reaching a higher level of production by changing the institutional structure and production technologies in the long run through the use of resources that are considered to be scarce in a country for production. Such growth has four basic factors: labor or labor force, natural resources, capital and technology level. Along with these basic factors, positive macroeconomic data that increase people's living standards such as purchasing power, employment, consumption, income and sales tax, increase in foreign investors are considered as growth indicators in an economy (Yetkiner, 2012: 8).

According to Amartya Sen, economic growth and development is the process of expanding the real freedoms that people enjoy. Sen sees the identification of economic development with narrower views such as growth in gross national product or increase in personal incomes or industrialization or technological progress or social modernization as contradictory when the focus is on human freedoms. Sen criticizes the fact that growth is measured in terms of per capita income and depends on market variables whose value can only be measured in terms of money, and that values such as the environment, human rights, social morality, unity and solidarity, which are not measured in monetary terms and do not have a label on them, are not included in the growth criteria (Sen, 2000: 3-4).

### 2.1.2. Determinants of Economic Growth

It is clear from the definitions of economic growth that the efficient use of labor, capital and technology used in production are the most fundamental factors of economic growth. From these sources, labor, which is also described as human capital, refers to the population that is employed or able to work in a country. It is observed that both production and consumption increase as the manpower that performs economically by participating in production increases (Yıldırım et al. 2013:501).

The period starting with Paul Romer's study on the determinants of growth in 1986 is characterized as the second period of the literature on growth. In this period, which is called the endogenous growth period or the new economic growth period, the factors that may lead to endogenous growth have been discussed in detail and many factors have been put forward. The most important factors of endogenous growth in growth models are technological change, human capital, education and public expenditures, as explained below (Yetkiner, 2012: 16):

1. Technology: Technology and technological developments, which are the most fundamental determinants of economic growth, are used to make the product more useful. In this way, it may be possible to increase the efficiency of labor and capital as production inputs and total productivity. However, alternative explanations are also available.

2. Human capital: As a reflection of the importance given to human beings in production models as well as in management, it is the general name of the skills that make the productivity of the worker evident with the education and health of the laborer. Human capital, which expresses the output of investments made in manpower, affects production efficiency.

3. Education: The effect of education on increasing labor productivity has led to the need for more and more educated labor force in production and has caused education to spread over a longer period of time over the years. Whereas labor was neglected in agricultural societies, in capitalist economies, an educated labor force, which means a qualified labor force, is more preferred. The health of the labor force, which makes labor continuity possible, is accepted as a component and even a prerequisite of human capital along with education. As a result, the approach that explains endogenous growth in terms of human capital argues that as the skills of the labor force increase, production efficiency increases, and therefore growth depends on the growth rate of human capital.

4. Public Expenditures: Public expenditures, which support production in the private sector, have a decisive role in endogenous growth. Public expenditures also ensure the production of goods produced within the scope of public investments and for public benefit, unlike private goods. Therefore, public expenditures are also considered as one of the determinants of economic growth.

Growth, whose main determinants in economic terms are technology, human capital, education and public expenditures, has many other determinants such as international trade, spillover effects, cultural factors, social capital and institutional structure.

### 2.1.3. Economic Growth Models

As explained in the first chapter, Smith, Richardo and Malthus have explained what economic growth is; but the models developed to measure growth emerged after the World War II.

Although the analysis of growth models is the subject of a separate study, the development of growth concepts and theories is summarized in Table 1.

*Table 1: Development of Economic Growth Concepts and Theories*

| Emergence of Growth Concepts and Theories | |
|---|---|
| Mercantilism | 15th Century |
| Physiocracy | Mid-18th century |
| Classical Theories | 1776 |
| Schumpeter's Theory of Innovative Growth | 1911 |
| Keynesian Theories | 1930s |
| Post Keynesian (Neo Keynesian) Theories | 1950s |
| Robert Solow's Neo Classical Theories and Exogenous Theory | 1950s - 1960s |
| Endogenous Growth Theories | 1980s - 1990s |

Source: Sharipov, 2015: 760.

This period, which started with Solow's article published in 1956 and lasted until 1986, is called the Neoclassical growth period. In this period when the first studies on the measurement of growth were carried out; it is known that it was measured with the models developed by Solow - Swan (1956)[11] and Ramsey depending on the variables of neoclassical economic theories. Although it was determined in the analyses of the neoclassical period that the main variable providing economic growth was technology, unlike capital accumulation, the existence of a theoretical relationship between technology and growth could not be demonstrated in growth models (Sharipov, 2015: 755).

### 2.1.4. Economic Growth Models

---

[11] Independently of Solow, Swan (1956), published in the same year, also describes the Solow model. For this reason, the Solow model is sometimes referred to as the Solow-Swan model.

As explained in the first chapter, Smith, Richardo and Malthus have explained what economic growth is; but the models developed to measure growth emerged after the World War II.

Although the analysis of growth models is the subject of a separate study, the development of growth concepts and theories is summarized in Table 1.

*Table 2: Development of Economic Growth Concepts and Theories*

| Emergence of Growth Concepts and Theories | |
|---|---|
| Mercantilism | 15th Century |
| Physiocracy | Mid-18th century |
| Classical Theories | 1776 |
| Schumpeter's Theory of Innovative Growth | 1911 |
| Keynesian Theories | 1930s |
| Post Keynesian (Neo Keynesian) Theories | 1950s |
| Robert Solow's Neo Classical Theories and Exogenous Theory | 1950s - 1960s |
| Endogenous Growth Theories | 1980s - 1990s |

Source: Sharipov, 2015: 760.

This period, which started with Solow's article published in 1956 and lasted until 1986, is called the Neoclassical growth period. In this period when the first studies on the measurement of growth were carried out; it is known that it was measured with the models developed by Solow - Swan (1956)[12] and Ramsey depending on the variables of neoclassical economic theories. Although it was determined in the analyses of the neoclassical period that the main variable providing economic growth was technology, unlike capital accumulation, the existence of a theoretical relationship between technology and growth could not be demonstrated in growth models (Sharipov, 2015: 755).

In a process in which economic growth is nurtured, influenced and transformed by many ongoing historical, cultural, economic and technological developments that has its beginnings in the period before World War I, the concept of worker and the working class were formed and women started to be employed as cheap labor. Unskilled workers in European cities that grew with industrialization consisted of women who since long ago had been involved in sectors such as agriculture, textiles and tobacco, as well as peasants who left behind their villages and came to the city, ending up being deprived of their lands and possessions. Due to the increasing mechanization and transformations in the new concepts of the city and the family that emerged as a result of the strengthening of a patriarchal capitalism, women were seen as a source of both unpaid and intensive labor in the home and cheap labor in the field of wage labor. Women were at the bottom of the wage pyramid. Due to their position at the bottom of the wage pyramid, women were employed more in factories, despite worsening living and working conditions. The number of women and children working under harsh conditions was so high that it is stated that the surplus value obtained from women and children was much higher than that of male workers, even though they received very low wages (Yılmaz, 2019: 130).

---

[12] Independently of Solow, Swan (1956), published in the same year, also describes the Solow model. For this reason, the Solow model is sometimes referred to as the Solow-Swan model.

This study focuses on the Solow and Ramsey models, which are neoclassical growth models that enable the interpretation and guidance of the changing conditions of economic life, and the Lucas model, which is one of the endogenous growth models due to its focus on population and labor input.

### 2.1.3.1. Solow Model

Solow's model defines two decision units: consumer families and producer firms. Consumer families, who own labor (L) and capital (K) inputs, earn income (Y) by renting these inputs to firms at time t, using technology A, for a fee. In Solow's economic growth model, income is shown as a function of labor, capital and technology (Ünsal, 2016: 112):

$$Y\mathrm{t} = F\,(K\mathrm{t}, L\mathrm{t}\,, A\mathrm{t})$$

It is stated that Solow's model, which enables the measurement of economic growth according to the increase in income, has made a significant contribution not only to economic growth but also to the adoption of the technology approach in macroeconomics (Acemoğlu, 2009: 26).

It is obvious that the supply-tended growth literature that followed Solow's model was criticized and found inadequate because it did not coincide with the changing behavior of the society due to the linearity in the hypothetical relationship between production and consumption, but it inspired different expansions and new models.

### 2.1.3.2. Ramsey Model

Criticisms of the linearity between output and consumption in neoclassical growth have led to a close examination of consumption behavior and the search for new answers. Before Solow's article, Ramsey's "Mathematical Theory of Saving", published in 1928, has defined saving as a determinant behavior that affects consumption and spreads over time. The demand-tended economic growth model, which is referred to as the Ramsey Model based on this finding of Ramsey and which was tried to be explained by Cass and Koopmans' applications in 1965, is known as the endogenous consumption-saving trade-off.

The model developed by Ramsey (1928), Cass (1965) and Koopmans (1965) is hypothetically free from all market imperfections and all problems arising from heterogeneous households and intergenerational linkages and operates in a natural benchmark situation. The lifetime utility function U defined for households in the Ramsey Model is as follows (Romer, 2012: 45):

$$U(c_t) = \int_0^\infty e^{-\rho t} \cdot u(c_t) \cdot L_t \cdot dt$$

Where $-\rho$ is the subjective discount rate[13], ct is the per capita consumption at time t, $u(ct)$ is the utility function of the person who consumes c at time t, and $Lt$ is the number of individuals in the household. The Model, which seeks to optimize the trade-off between consumption and saving, has assumptions that take into account the individuals forming the household (Romer, 2012: 46-51):

- Households grow at rate $n$ and are numerous and identical.
- All individuals are members of the household and supply labor per unit of time.
- Households rent their factors of production to firms.
- Household income consists of the income from the factors rented to firms.

It is also stated that the endogenization of the consumption-savings trade-off, which Solow regards as an exogenous value, in the Ramsey Model, which is actually a Solovian model, cannot explain growth. However, the Ramsey Model allows for the analysis of the amount of income required to ensure the welfare of households and to maintain it throughout life; as well as decisions on consumption - saving and the optimal control theory, and different variables (Yetkiner, 2012: 85-126).

The Ramsey Model approach used to explain the utility maximization of households makes the observer think that all individuals are obliged to firms. Therefore, as mentioned before, everything seems to be as it should be for the firms established by the owners of capital, who leave the state out of the game and determine the market rules to ensure that all resources are in their own use. Therefore, models that prioritize the profit maximization of the owners of capital over that of households cannot explain how growth occurs in the long run. The absence of an explanation for the labor of housewives, which we know is not hired labor; is an indication that the main focus of the Model is not on households, as it is understood from the assumption that all individuals constituting the household supply labor at every point in time.

---

[13]Since the variable compared over time in the utility function is not a monetary product with an interest rate, but a benefit, the subjective comparison or proportioning of the benefits obtained at different times is made with the $\rho$ coefficient (Yetkiner, 2012: 96).

David Romer (2012) argues that the Ramsey-Cass-Koopmans and Diamond models are similar to the Solow model, but the dynamics of economic aggregates are determined by decisions at the microeconomic level. The main difference between the two models is that the Diamond (1965) model assumes the continuous entry of new households into the economy. Both models continue to take the growth rates of labor and information as given, but as a result of deriving the evolution of the capital stock from the interaction of maximizing households and firms in competitive markets, it is explained that the saving rate is no longer exogenous and need not be constant (Romer, 2012: 49).

Endogenous growth models, on the other hand, are also known as linear growth models and Y = AK models since only technology (A) and capital (K) are taken into account when calculating output. Due to their simplicity, linear growth models assume that positive population growth is necessary for long-term growth and that the long-term growth rate of the economy increases with population growth. That is, long-term growth is shown to be an increasing function of the population level. However, the growth of labor is considered exogenous in this model (Romer, 2012: 101-125).

### 2.1.3.3. Lucas Model

It can be observed that the developments considered as a new milestone in economic growth theory started in the 80s-90s; and in these new growth theories, for the first time, American economists Paul Romer and Robert Lucas developed hypotheses about the endogenous nature of technological innovations based on technological development and human capital. Contrary to neoclassical theories, endogenous growth theories are in favor of state intervention in the development process, and the first[14] group of these theories, which are divided into two groups, consists of the theories of P. Romer (1989) and R. Lucas (1988), in which human capital is considered an important determinant of economic growth. The key factor in Paul Romer's endogenous growth theory is the "knowledge" variable. In the theory, which assumes that knowledge is accessible to everyone and can be used at the same time, Romer assumes that the total amount of human capital remains constant throughout the time interval considered (Sharipov, 2015: 769).

---

[14]In the second group of theories, J. Grossman and E. Helpman's theory explains R&D as a key factor of growth in the impact of endogenous high-tech innovations on economic growth rates.

Lucas' theory, in contrast to Romer's theory, is based on human capital accumulation. This is a powerful process of accumulation using specific resources, which may incur alternative costs. Lucas argues that people may want to spend their time either participating in existing production or accumulating human capital. According to Lucas, it is the allocation of time between these alternative paths that essentially determines the rate of economic growth. For example, even if a reduction in the time spent on production naturally leads to a reduction in the quantity of output, accelerated investment in human resources can increase the growth of output. Therefore, the difference of this theory is that it focuses on human capital, which is included in the production function together with the education factor (Sharipov, 2015: 771).

The Lucas model is considered as a model in which the exterior effect of labor, i.e. human capital, can be questioned. In the Lucas model, where the defined firm is assumed to produce for a single household, it is assumed that the household consists of two people, the "producer" and the "shopper"; and the communication between them is limited. Lucas makes a simplifying assumption regarding the producer's production decision; he assumes that the producer will produce as long as the price he offers meets the customer's expectation. While this behavior of mutual certainty is not the same as maximizing expected utility, it is generally observed that the utility-maximizing choice depends not only on the household's preferred price but also on its uncertainty. Lucas's inclusion of rational expectations in macroeconomics, although initially quite controversial, is now recognized as one of the strong implications of his model. According to this assumption, assuming that the producer chooses how much to produce according to the mathematical expectation of what the price set by the customer will be, is implicitly found to be the rational expectations of the producer. Today, this assumption of rational expectation is considered as normal as the assumption that individuals maximize the utility (Romer, 2012: 290-295).

## 2.2. GDP AS A GROWTH INDICATOR and ITS CALCULATION

In order for an increase in economic activity and per capita output to be defined as growth, these increases must be realized in the long run through more efficient use of production potential. For this reason, the problem of economic growth is generally seen as a long-run problem. Development, on the other hand, can be explained as the rise of people at a certain level of welfare to a higher level of welfare, including economic growth, and the balanced increase in productivity of all segments of activity.

Due to their position at the bottom of the wage pyramid, women were employed more in factories, despite worsening living and working conditions. The number of women and children working under harsh conditions was so high that it is stated that the surplus value obtained through women and children was much higher than that of male workers, even though they received very low wages (Yılmaz, 2019: 130).

This study focuses on the Solow and Ramsey models, which are neoclassical growth models that enable the interpretation and guidance of the changing conditions of economic life, and the Lucas model, which is one of the endogenous growth models due to its focus on population and labor input.

### 2.2.1. Definition and Importance of GDP

GDP is an economic quantity that measures the total value of all final goods and services produced within a country's borders over a given period of time. It should not be confused with Gross National Product (GNP), where the value of goods and services produced in a country by its citizens is calculated regardless of whether they live inside or outside the country. GDP is used by economists to determine the performance of the economy and economic growth, as well as to measure the effects of deflation and inflation.

There are four different types of GDP - real, nominal, actual and potential - and it is important to know the difference between them, as each represents a different economic outlook. Real GDP is an inflation-adjusted measure of GDP that uses fixed prices for goods and services, usually based on the price levels of a predetermined base year or the previous year; and is considered to be the most accurate depiction of a country's economy and economic growth rate. Nominal GDP is the inflation-adjusted measure of the prices of goods and services at current price levels. Real GDP is the size of a country's economy measured at the ~~current~~, current time. Potential GDP refers to the size of a country's economy when calculated under ideal conditions such as a stable currency, low inflation and full employment (Krugman, 2021: 5).

A rising GDP in a country means that incomes are rising and consumers are able to meet all their needs by buying more, which is important because it shows the strength of the economy. However, GDP alone cannot show the standard of living as a whole. China, for example, despite its large GDP, has a low standard of living and is classified as a middle-income country. GDP growth in consecutive quarters over a year indicates that the economy is also expanding, while a negative GDP growth rate in two or more consecutive quarters is considered a recession. These changes suggest to economists and policymakers that measures to boost economic activity are necessary to maintain stability. GDP data is also considered important as it is an indicator of the growth changes of all countries or economic regions of the world relative to each other (Krugman, 2021: 8).

### 2.2.2. Calculation of GDP

GDP is one of the most widely used tools for measuring a country's economy and is calculated by countries themselves, as well as sometimes by world organizations such as the World Bank, the International Monetary Fund and the United Nations. There are three different ways in which a country's GDP is calculated by economists and statisticians, and theoretically they are all expected to ultimately produce the same number.

GDP can be calculated by calculating the production carried out within a country and the market value of that production. GDP can also be calculated indirectly by using total expenditure as the value of everything purchased within a country and the value of that country's net exports to other countries. As another indirect way of calculation; the sum of the incomes of real and legal persons in the country constitutes the GDP (Ünsal, 2013: 45).

### 2.2.3.1. Production Method

When calculating production-based GDP, the value added of all domestic production is taken into account. Value added is the price at which a seller sells a product minus the price at which the seller buys the product from the supplier. In this method, the monetary value of all final goods and services produced in an economy for a year is calculated as the flow of goods and services. Final goods here refer to goods that are directly consumed and not used in the subsequent production process.

Goods that are used further in the production process are called intermediate goods, and since the value of intermediate goods is already included in the value of final goods, the value of intermediate goods is not included in national income calculations. Otherwise, the value of goods is calculated twice and a double counting problem arises. To avoid this problem, the value-added method is used, in which not the entire value of a commodity is calculated, but the value added at each stage of production (i.e. the value of the intermediate good to the final good) and summed up. GDP is calculated directly from the value added of all firms operating in the market (Ünsal, 2013:46):

GDP = Sum of Value Added of All Firms

In Turkey, according to TurkStat, GDP is calculated by the production method and is the sum of the value of all goods and services created by the productive units in an economy as a result of their economic activities in a certain period, minus the sum of the inputs used in the production of these goods and services (TurkStat, 2020)

### 2.2.3.2. Expenditure Method

Keynes's expenditure method theory of national income accounts is still widely accepted. The expenditure method is based on the idea that all final goods and services produced in an economy must be purchased by someone. Goods that are not sold are accounted for as purchased by the producer. To better understand what GDP is and why the number is important to economists, each component of the calculation, known as consumption, investment, government spending and net exports, is considered important for a demand-tended economy.

Consumption expenditure (C) is the largest part of the GDP calculation and includes everything that requires a household or individual to spend money, such as non-durable goods like food, clothing, etc. and rent. The purchaser and the consumer may not be the same person, nor is it possible for everything purchased to be consumed in the same GDP period. Investment expenditure (I) is typically defined as investment in new materials and equipment by an enterprise for production purposes; for example, an enterprise buying new computers for all its employees or households buying properties which are included in investments, since houses are bought on credit (most households do not pay all the cash upfront for the property). Investments such as the purchase of shares are not included in this definition of investment as they are considered savings. Government expenditure (G) is the sum of everything the government buys or spends. This means any physical goods purchased by the government, such as fire trucks or aircraft carriers, investments made by the government, and salaries of government employees; like teachers for example. Exports consist of all goods produced in the country and sold to other countries; imports consist of all goods produced in other countries and imported into the country. To calculate net exports (X-M), the amount of imports (M) must be subtracted from the sum of exports (X). **Using** the expenditure method, GDP is calculated as follows (Krugman, 2021: 5):

GDP = C + I + G + (X - M)

As seen in the formula, in this method, national income is measured as a flow of expenditures. Expenditures are consumption (C), investment (I), government expenditures (G), exports (X) and imports (M); GDP is the sum of private consumption expenditures, gross capital formation (government and private), government consumption expenditures and net exports (exports-imports) (Ünsal, 2013: 46-48).

### 2.2.3.3. Income Method

In his analysis of the General Theory of Employment written in 1936, Keynes states that the propensity to consume, the marginal efficiency of capital and the interest rate as independent variables depend**ing** on the volume of employment and national income, calculated in terms of wage income. However, while determining the distribution of national income, he states that they do not take into account the effects and consequences of the changes caused by the labor element in question (the available skill and quantity, the quality and quantity of equipment, the available technique, the degree of competition, the tastes and habits of the consumer, different labor intensity and activities, and factors of the social structure such as control and organization) (Keynes, 2018: 215-218).

In measuring economic growth by income growth, Romer (2012) argues that studies reporting and discussing basic data on average, real incomes in modern history have uncertainties about the extent of long-term growth; most of which are related not to the behavior of nominal income but to the price indices needed to convert these figures into real income estimates. While Romer notes that adjusting quality changes and introducing new goods is conceptually and practically difficult, there are arguments that traditional price indices are biased and do not do a good job of making these adjustments (Romer, 2012: 6).

As components of income-based GDP calculations, income is defined in various ways through four factors of production. These are: Labor and wages for laborers, rent from land rental, interest return on capital, and entrepreneurs' profits. The calculations take into account workers' salaries, gross surplus and mixed income, as well as taxes and subsidies. According to these explanations, the following formula is used to calculate income-based GDP (Krugman, 2021: 9):

GDP = Employee salaries + Gross operating surplus + Gross mixed income + (Taxes - Production and Import Subsidies)

Employees' salaries are the sum of payments made to all employees or workers, including social benefit payments such as social security. Gross operating surplus is included as the profit of large enterprises with the status of joint stock companies with a large number of employees. Profit of unincorporated enterprises is included as gross mixed income, which consists of the income of self-employed individuals such as doctors, lawyers, etc. who employ their own labor and capital. According to these explanations, the following formula is obtained when calculating income-based GDP (Ünsal, 2013: 52):

GDP = Labor Income (wages and salaries) + Capital Income (Rent, Interest, Net Profit) + Indirect Taxes + Depreciation

The income approach essentially reflects not the purpose for which the final goods produced in a country in a year are used, but the shares of the factors of production in the functional income distribution, i.e. the shares of those who participate in the production process in a given country around the year (Ünsal, 2013: 53).

In Turkey, according to TurkStat, GDP is calculated by the income method, using salaries, wage income, business profits and various tax revenues obtained by the state; what is taken in to account is the sum of the values paid to the factors of production by the productive units involved in the production process of goods and services in each branch of activity (TurkStat, 2020).

## 2.3. HOUSEWIFE LABOR IN ECONOMIC GROWTH and GDP CALCULATIONS

While the participation and employment of women as-paid but cheap workers in the labor force has been increasing since the 1950s, after World War II, the issue of whether the labor of housewives should be taken into account has started to be evaluated with the paradigm shifts that emerged in heterodox theories, as explained in the first part of the study.

In the modeling of economic growth, the Lucas model is the closest to the distribution of labor within households in terms of labor productivity and efficiency. In models that try to explain growth, it is thought that the reason why housewives' labor is not included among the determining factors in human capital as one of the main sources of economic growth is; showing similarity with the reasoning for exclusion of physical capital in long-run calculations; the focus is on short-term capital accumulation and the cost of physical wear and tear in the long run is passivized. Although this situation seems to be consistent with an economic approach that prioritizes the interests of capital owners before the interests of the society as a whole, it shows that in the existing growth models of the literature on economic growth, which is closely related to the labor structure as an important issue of macroeconomics, the relationship between labor-employment-growth has not yet been fully explained by considering the interests of all parties. When we move from income calculations to distributional calculations in growth models, the fact that housewives' labor is stated as a non-employment labor factor and not allocated an allowance, despite the fact that it is implicitly included in households, suggests that all calculations regarding distribution are incomplete or erroneous.

The problems associated with estimating real output have led official government statistics to underestimate the growth rates of real GDP, real personal income and productivity, Feldstein argues, and this underestimation is important not only for economists trying to understand where the economy is heading, but also for the wider public and politicians. The understatement of real growth reflects the enormous difficulty of dealing with quality change and the even greater difficulty of measuring the value created by the introduction of new goods and services. Feldstein notes how flawed official estimates actually are, despite the huge amount of attention devoted to this issue in the economic literature and by government agencies (Feldstein, 2017: 3).

### 2.3.1. Family Labor Demand and Housewife's Unpaid Labor Supply

Contrary to popular belief, it is not always employers who demand labor, nor is it always workers who supply labor. Labor is the demand of human beings, the most prominent subject of an economy that has been developing for centuries to meet their needs; and this demand begins in the family. We learn to demand from our families, primarily from our mothers. For this reason, the effort to make visible the labor that has been compressed within households is shown to make economic science more useful and coherent.

In general, growth models that consider the female labor force as wage labor, calculate productivity in terms of the skilled nature of labor, and women are characterized as low-productivity and unskilled workers due to their lack of education and limited working hours. In GDP accounts, as in economic growth accounts, in addition to women working with low wages in employment, the labor of housewives, who supply labor in response to the demand of the family, is not shown or accepted as of any quality.

In terms of the labor market, women's domestic production does not conform to the characteristics of a perfectly competitive market. Because the woman working inside the house (the mother of the house) cannot produce wherever she wants, as much as she wants and whatever she wants; she creates an economy within the limited conditions of the house and only for the needs of that household.

According to Veblen, consumption is considered as the end of acquisition and accumulation. The fact that consumption takes place by the owner or by the household, which in theory is identical with the labor equivalent, is characterized as the legitimate end of acquisition. This legitimacy should be taken into account in economic theory as a natural consequence of the traditional acquisition by the consumer who owns goods with supposedly higher physical wants - not for secondary needs such as aesthetic, intellectual or different service designs (Veblen, 1918: 25). With this approach, Veblen emphasizes that consumption is an economic value rather than being the subject of where and how production takes place.

Other unreported forms of labor are defined as labor that is not included in GDP, such as childcare, cooking or cleaning, i.e time for labor which is mostly spent by housewives. This means, for example, that the sector of the economy made up of men and women who choose to stay at home in order to care for their children while working full-time, is not included in national accounts of the economy or labor force. While GDP is a very important measure of economies, it is cited as the reason why economics is an important but imperfect science, because it does not take into account the general well-being, health and happiness of the population (Krugman, 2021: 5).

### 2.3.2. Labor-Employment Dependency and the Invisible Labor of Housewives

In the first part of the study, it was explained that classical economists evaluate labor supply and demand in relation to the real wage; Keynesian economists evaluate labor supply in relation to the nominal wage and labor demand in relation to the real wage, and Monetarists evaluate labor supply in relation to the expected real wage and labor demand in relation to the actual wage. All these considerations suggest that the dependence between labor and employment is one of the reasons why the labor of housewives is not taken into account in economic calculations. Table 2 shows labor supply and demand as accepted in scientific theories. When examined carefully, it actually shows the labor-employment dependency and the neglect of housewives' labor in this dependency due to its unpaid nature. For this reason, the Table 2 is named here as "Labor-Employment Dependency and Invisible Housewife Labor", different from the name in the source from which it is taken.

*Table 3: Labor-Employment Dependency and Invisible Housewife Labor*

| | Labor Supply | Labor Demand |
|---|---|---|
| **Classics Klasikler** | **Real Wage** $L_S = f(\frac{W^{(+)}}{P})$ $\frac{W}{P} \uparrow$ $L_S \downarrow$ (Vice versa) | Real Wage |
| **Keynesians** | **Nominal Wage** $L_S = f(W^{(+)})$ $W \uparrow$ $L_S \uparrow$ (Vice versa) | $L_D = f(\frac{W^{(-)}}{P})$ $\frac{W}{P} \uparrow$ $L_D \downarrow$ |
| **Monetarists** | **Expected Wage** $L_S = f(\frac{W^{(+)}}{P^e})$ $\frac{W}{P^e} \uparrow$ $L_S \uparrow$ (Vice versa) | $\frac{W}{P} \downarrow$ $L_D \uparrow$ |

Source: Bilgili, 2015: 65 (Labor Supply and Demand)

In Table 2, LS is the labor supply of the labor market, $L_D$ is the labor demand, W is the nominal wage paid to the worker, P is the price of the product, and $\frac{W}{P}$ is the real wage level. Since labor demand and labor supply are considered as a function of the real wage, for example, the Classical notation $\frac{W^{(+)}}{P}$ shows that labor supply is an increasing function of the real wage while $\frac{W^{(-)}}{P}$ labor demand is a decreasing function of the real wage (Snowdon and Vane, 2005: 32-36).

It is stated that the reason for the convergence of less developed economies to developed countries in national income accounts, is the consideration of the purchasing power of output (PPP) in expenditures. This is because the production of unemployed housewives at home and in the fields, included in these expenditure calculations, does not take into consideration the income they earn from this production and consequently are not included in national income calculations. Therefore, the labor of housewives which is not included in employment is an important component of the agricultural and service bound sectors (Aytekin, 2017: 163).

Associating evaluations about the place of labor supply and demand in economic theories only with wages and employment, means associating social welfare only with money. Considering the responsibility and impartiality of economics in developing appropriate policies beyond measuring only the monetary equivalent of social welfare, it also has the responsibility to determine regulatory policies regarding the distribution of all inputs. Therefore, there is a need for scientific truths to develop the necessary policies on the valuation of housewives' labor. Evaluating the supply and demand for labor only in the interests of the owner of capital, seems to contradict the impartiality of science.

After the theoretical analysis so far, since the last part of the study will attempt to measure the effect of housewives' labor on GDP, it will be useful to conceptualize our main variable- housewives' labor- at this stage.

## 2.4. CONCEPTUALIZATION OF HOUSEWIVES' LABOR

In economics, it is accepted that subsistence depends on labor power on the basis of meeting human needs. While in the pre-capitalist period, the concept of labor was accepted as a natural part of life and thus of economic functioning for both men and women. However, in the period of capitalism, under the influence of the demand determined by the owners of money, a differentiation is made on whether the owner of the labor is a male or a female.

As a remnant of the culture of obedience, created for peasants by the feudalism of the Middle Ages, when the class structure of society was formed and inherited by the capitalism that developed afterwards, it is seen that women are far from showing a voluntary attitude towards their labor power. In the conceptual construct of capitalism, the demand for labor determines the laborer's ability to find a place in the labor market.

The fact that the labor of housewives, as described in orthodox economic theories, is not included in the calculations of economic theories does not mean that such a labor power does not exist; nor does it mean that it is not a real demand. The demand for the labor power of housewives is indispensable and constantly present in the economic structure of social life. This difference between fiction and reality appears as a discrimination against women. At this stage, conceptualizing the labor of housewives is seen as an important necessity.

In this section of the study, it will be revealed, how the labor of housewives, who are excluded from employment becomes ambiguous within the gendered division of labor and clarified through the "labor equality" approach. Subsequently the "bounded rationality" will be taken into account as stated in the heterodox economics literature, to define the concept of housewives' labor.

### 2.4.1. Labor Equality Disappearing with Gendered Division of Labor

Although women's labor has been an economic force for the development of society since the earliest ages of history, their gender has been the cause of discrimination in all spheres of political, religious and economic life, which has progressively diminished the role of women.

According to Friedrich Engels, Paleolithic and Neolithic[15] artifacts from early human societies show a small and sharply demarcated division of labor based on sharing and cooperation. In this division of labor, all work done by men and women was considered valuable and individuals of both sexes gained respect by improving their work. Hunting groups of non-motherly women systematically hunted and sometimes fought. The division of labor did not make women sedentary and immobile. It is stated that there was no inequality in this division of labor, in which women with children took over the work around the house and were mostly engaged in subsistence production based on agriculture and textiles, while men were engaged in herding and hunting (Engels, 2012: 26).

In Veblen's works, it is stated that the painful process of transition from the matriarchal period to the patriarchal period started in the barbaric stages of the subculture when women were captured as booty in wars due to their matriarchal assets. It is seen that slavery, which emerged with the spread of slavery to other women and captives in addition to the women captured from the enemy, continues in some cases by transforming into a property marriage in which the woman, who is transferred to the protection of the man as a property, is passivized and the head is a man. The tradition of property created by the predatory culture transforms slavery and marriage into a social status decided by men. According to Veblen, as long as the male-headed household exists, it proves that the woman is a servant in economic terms and maintains a relationship of submission. Women are expected to fulfill their prescribed duties and to be perfect in obedience, equivalent to that of a slave and a housewife equipped with skills and abilities compatible with this basic benefit they provide (Veblen, 1918: 30-50).

---

[15]The Paleolithic (Chipped Stone) Age began about 2 million years ago and ended 10,000 years ago. The Neolithic (Polished Stone) Age is the period between 8000 - 5500 BC.

In the 17th and 18th centuries, despite the transition from a feudal economy to an industrialized economy, it is seen that the return of women's assets brought as dowry[16] through marriage contracts, under the pressure of the church, an influential institution of the period, led capitalists to "invest in women, not in land". Despite this idea that rendered women passive by turning them into an instrument of commerce, the French Revolution (1789) transformed marriage into a civil contract and the family, which represented the private sphere, became able to direct the political society, providing women with the opportunity for freedom and equality. However, the separation of private life from public life also separates civil life from political life and deprives women of the opportunity to participate in decisions. In addition, loyalty and responsibility to the family also limits women in civil life. In the Civil Code adopted in France in 1804, the statement "*The sole and absolute judge of family honor is the man*" indicates that the principle of inequality is valid, not gender equality.... It is stated that with the French Civil Code, which spread rapidly in Europe and was taken as an example by many countries around the world, women's struggle not only against the gender-based division of labor in society, but also against the inequality reinforced by laws began (Tanilli, 2006: 38).

In an interview Friedman gave to a newspaper in 1970, he stated that a 'businessman' who is self-elected or appointed directly or indirectly by shareholders should be a legislator, an executive and a jurist at the same time. He or she would decide who should be taxed how much and for what purpose; and these taxes to rein in inflation, improve the environment, fight poverty, etc., would be spend and guided only by general advice from on high (Friedman, 1970: 4).

---

[16]It is stated that the property under the control of women in the matriarchal period is defined as dowry with the concept of property rights that emerged with the transition to the patriarchal period.

Considering these explanations for entrepreneurs, known as the Friedman doctrine, it is clear that the person holding all this power, is of course a man, as the term 'businessman' implies, and it also shows how patriarchal the economic system became in the 1970s, even as it developed. In this case, the man is the one who decides everything in important matters, what the man owns is important and valuable, the man is the owner of both money and labor, the owner of labor earns money, and for these reasons labor is very important because it represents masculine power.

It is seen that the division of labor according to gender is based on the necessity for women to be close to home due to their responsibilities related to childbearing and childcare. With the development of civilization, such requirements are manifested as a deterioration in labor equality and "gender inequality" as it is known in the literature.

### 2.4.2. Limited Rationality and Housewife Labor

As explained within the heterodox schools of economics, Simon, the founder of behavioral economics, emphasizes that individuals interact with institutions and social structures in a reciprocal way; and when considered for all individuals that make up the society, they form a collective consciousness. In this case, the individual who cannot act independently in his/her decisions and who bears a responsibility towards the society in which he/she resides is defined as a "bounded rational individual" (Simon, 1997: 96).

Limited rational individuals, when implementing the task or decision expressed in general, are said to replace the global rationality of economic man with the type of rational behavior that is compatible with access to the information and computational capacities that they actually possess in environments where similar organisms like themselves exist (Wheeler, 2020, 1)

It is known that the vitality that enables the emergence of human labor as human capital begins with the birth of human beings into this world and that they are in need of meeting their basic needs until they are able to function. This neediness limits the rational behavior of the housewife and transforms her into a limited individual who prefers to fulfill the responsibilities arising from the expectations of her child and family members. In this study, the housewife is defined as a limited rational individual.

In contrast to the rational behavior of individuals defined as 'homo economicus' and the interest groups they form by pursuing their own interests, limited rationality refers to the behavior of the individual by pursuing the interests of those who are related or responsible for him/her as well as himself/herself. In this study, the rationality of houskeeping behavior, which is the method preferred by housewives, who most closely fit the definition of bounded rationality, while performing labor supply, is rated in Table 3.

Table 3 is inspired by the 5-point Likert scale, a widely used behavioral scale developed by R. Likert in 1932. In the ranking of rationality behavior on a scale of 1-10, 10 being the degree of Rationality and 1 being the degree of Limited rationality, an even number is used for each degree in five different interest groups identified, as there is often more than one type of profile. The effort made by housewives, ~~which is~~ being considered outside the labor-employment dependency in the economic system, is different from a labor supply in return for interest; it is realized at the highest level of limited rationality, which is satisfied with the returns provided by the family of which she is a member. As stated in the table, housewives are considered to have the lowest rationality when it comes to an action of her own interest.

*Tablo 4: Degree of Rationality and Limited Rationality of Interest Groups*

| No | Groups | Interests to be pursued in labor supply | Rationality | Rationality Limited |
|---|---|---|---|---|
| 1 | Managers, Decision makers | Staying in management | 10 | 2 |
| 2 | Merchants Intermediaries | Profitability | 8 | 4 |
| 3 | Labor owners, Producers | Adapt to market conditions to increase their earnings | 6 | 6 |
| 4 | Households, Consumers | Act according to their income | 4 | 8 |
| 5 | Housewives | Fulfilling what is given to them | 2 | 10 |

Source: Prepared by the author of the study.

Under certain conditions, production, which is at the center of economic laws laid down by induction and controlled by deduction, is defined as the creation of utility, which is defined as the capacity of individuals to satisfy their desires. Increasing the production of the labor producer, defined as the effort made by a person for the realization of production, i.e. increasing the labor supply, cannot be realized whtin a short period of time, since it takes about twenty years to train an employee (Tolfree, 1974: 5-7). Therefore, it should not be ignored that the labor of housewives is the main actor in increasing the labor supply as labor input.

In this study, the housewife, who is the carrier of vitality that ensures the continuity of the generation and the provider of the basic needs of her baby, is defined as a bounded rational individual, while housewife labor is considered as the basic labor force that ensures this continuity. In this case, when the labor of a housewife is evaluated within the understanding of bounded rationality, it is revealed that it is the labor of an individual who prefers to work at home to provide for her family. Based on the people-oriented approach of economics, "housewife labor" is defined as the labor that women use especially and primarily for all activities related to housework and motherhood, which is currently unpaid but creates an economic value.

## CHAPTER THREE
## 3.THE EFFECT OF HOUSEWIVES' LABOR ON GDP ACCOUNTS, THE CASE OF TURKEY

The rapid population growth that started in the late 1800s has increased the interest in the science of economics; and still with the developments in many fields, the need to examine the activities that concern human beings and their needs, not only from the micro scale of economics but also from the macro scale has turned out to be a necessity. Considering that the world population has increased by more than six billion in the last two centuries, it is necessary to include disciplines other than economics in the solution while evaluating economic problems on a macro scale. This necessity also includes the expectations of other disciplines seeking solutions, regarding the responsibility of economics to solve the problem of distribution.

One of the most fundamental problems of economics, the problems of production and distribution, is the confusion between the place of the human being in the production processes carried out to meet his/her own needs and the uncertainty of the place of the state that emerges in these processes. When distribution is defined as the ability of those who have contributed to the creation of the whole to be compensated for their labor, it becomes important whether it is done fairly or not. Therefore, in order to solve the problem of distribution, first of all, it is necessary to determine the whole to be divided correctly and to know where the responsibility of the state to take care of its citizens begins and ends.

The production and distribution mechanisms commonly used in the economic sense are known as either capitalism, where the owners of capital are the determinants, or socialism, where the central authority is the determinant. In capitalism, the problem of distribution is defined as the conflict between the components commodified as factors of production, while in socialism, it is stated that workers are in direct conflict with the state as the central authority. In both systems, the passivity of factor incomes, which is determined by the state or capital, leads to exploitation. While exploitation manifests itself in socialism by trying to ensure equality in proportion to the qualifications of the individual, in capitalism it is seen directly as income inequality (Bocutoğlu, 2012: 9).

Meghnad Desai, in his book "Marxist Economic Theory" published in 1974, emphasizes that Marxist and Ricardian approaches are largely similar on methodological grounds and states that the crucial role in the theory of value for both is not in exchange relations but in production relations. Moreover, both approaches recognize the existence of competitive forces that seek to equalize the rate of profit across the economy. According to Desai, although the concept of "value" is essentially philosophical in nature, the adequacy of a theory of value is important not only in terms of dealing with theoretical problems but also in terms of answering practical and operational questions (Desai, 2009: 6-7).

The main questions that need to be answered in this research are the accuracy of the inputs taken into account in GDP calculations and how the data omitted in the calculations affect production and distribution problems. In seeking answers to these questions, it is important to note GDP calculations, as an important measure of economic welfare;

1. Determining the amount of labor,

2. One reason for distributional problems is that the unpaid labor of housewives is not economically accounted for,

3. How GDP would be affected by taking into account the labor of housewives,

4. It is thought that there will be progress in efforts to resolve distributional problems.

The daily work known as housewives' work is undoubtedly more numerous and more noteworthy than the differences in working hours of the employed worker, which must be carefully calculated because they are paid for by the employer. However, it is not possible to find the economic equivalent of housewives' labor in the concepts of employment and labor in the 21st century, when women's rights are still being defended to eliminate gender inequality and ensure equality, perhaps because no one undertakes to pay for it.

The contribution of the study to the Labor Theory should be sought at this point. In fact, development is based on economic growth, economic growth is based on GDP, and GDP is based on production-income-expenditure calculations. Therefore, in order for the calculations to be made correctly, it is necessary to define the labor element in the economic sense not only as dependent on the condition of employment, but also in accordance with human dignity and reason.

## 3.1. LITERATURE REVIEW

In this study, which focuses on contributing to the definition of labor by providing economic visibility to the labor of housewives, the theoretical literature on labor is presented in the first section. While trying to include the literature on empirical studies here, a review of studies that deal with the measurement of housewives' labor outside the employment condition and/or take into account housewives' labor in GDP calculations has been attempted, although the studies are insufficient.

According to Veblen, who argued that rationality is an obstacle to human economic development, men are preferred in important jobs because they are more suitable for the job due to their greater time and human energy. Therefore, in the economy of the leisure class, it can be observed that under the influence of industrialization, the busy housewife of the early patriarchal days was replaced by the lady[17]-in-waiting and the servants. Veblen thinks that at all levels and walks of life and at any stage of economic development, the leisure of the lady and the servant differs from the leisure of the gentleman in that it is apparently a laborious occupation. The tasks performed by the wife or domestic servants are usually sufficiently arduous and at the same time directed towards ends that are considered extremely necessary for the well-being of the entire household. Insofar as these services contribute to the physical efficiency or comfort of the master or the rest of the household, they should be considered productive work. Only after deducting this productive work should a certain amount of time be identified and classified as leisure performance, since, as with any work, the residue of work is clearly wasted. This is because, according to Veblen, these domestic occupations have the advantage of properly asserting the grounds of the private economy, essentially as a method of attributing material prestige to the master or household (Veblen, 1918: 57-60).

---

[17]By definition, a lady is a woman of superior social standing, especially a noblewoman.

In his study, Meçik (2021) states that the most important deficiency in the Turkish labor market is that although the supply and demand relations and mismatch problems in the market are known, the connections of the labor market with other areas are not taken into account in a holistic manner. According to Meçik, the important consequences of the social approach in the established culture, which translates into gender inequality in general, cannot produce solutions to ensure employment balance in the market. It is stated that this problem creates a negative input of macroeconomics in policies related to labor markets that lasts from generation to generation (Meçik, 2021: 128).

In his study with data from the World War II period and its aftermath, Perry reveals that as a result of the employment of more women and young workers, the growth rate increased by 0.2 percentage points, between 1945-1965 and 0.5 percentage points between 1965-1970. The reason for this increase is stated to be the increase in male employment again after the war as an employment variable (Perry, 1971: 537). The measurements in Perry's study prove two aspects at once: that women workers in employment were paid relatively lower wages; and that women worked in employment as well as housework despite the war conditions. Perry does not question about 'who did people's domestic work' during the war period; nor does he make any reference to the fact that an unusual number of housewives, who normally do not work at all, contributed to 0.2 percentage points of growth by working both at home and outside the home, despite being seen as weak workers.

Bryant and Zick, in their 1990 and 2005 studies on the economics of women's employment and fertility, show the fertility implications of women's increased participation in the labor force and point to the decline in the labor force in the long run. While the labor force participation rate of housewives, which was 4.5% in the 1900s, has increased revolutionarily in a hundred years, their fertility has decreased. In a study conducted in the US, the birth rate dropped from 55.2 births per 1000 population in 1820 to 32.2 in 1900 and 13.9 in 2002 (Bryant and Zick, 2010: 125-198). Bryant and Zick's study on the economic organization of households is important in that it identifies, that men's working hours have shortened in proportion to the increase in women's employment, although it does not economically measure the labor of housewives.

According to Myles and Quadagno, in order to maintain high levels of employment in developed countries, despite the aging of the population and the decline in male labor force employment with the population, resources in the welfare state should be used in a way to maximize the employment of women. Myles and Quadagno argue that in welfare states, a financing policy can be followed according to the contribution of women's increased labor force participation to national income on the one hand and their willingness to reproduce the next generation on the other (Myles and Quadagno, 2002: 41-42). However, it does not suggest how the calculations related to this issue should be made.

There are many examples of gender inequality in the literature of economy, where the owner and winner of labor power is a male. It can be observed that studies on the effects of gender equality on development are also based on increasing women's employment and conclude that an increase in the number of women employees positively affects development. In fact, increasing the number of employees already positively affects development. It is explained in the classical labor theory section, why employers prefer male workers is, that the gender of the employees being female does not have a positive effect on development. In addition, many theories of labor to date do not deal with the political economy of whether the owner of labor is a woman or not. At a time when the science of economics was still trying to establish itself, it is seen that not only the labor of housewives, but even women who actually worked as workers in factories were engaged in an uphill struggle for fair wages and working conditions. For example, in a study conducted by Frager and Patrias in 2005, it is stated that the years between 1870 and 1939 were important as a period of growth of industrial capitalism in Canada and also a period when many women joined the paid labor force. The study reveals that women in employment, whether in casual, salaried or low-status, unskilled or managerial positions, were limited by earnings inconsistencies and jobs that were seen as less valuable than men's jobs (Frager and Patrias, 2005). This study by Frager and Patrias also shows that capitalist priorities prevail and the focus is on the problems of women in employment rather than on the labor of housewives. However, as mentioned earlier, such studies are important in understanding that the reason why labor theories are not interested in women is that they were not already in the labor market when the concepts that regulate working life (wages, working hours, unionization, etc.) emerged and were determined according to men, and that there was no definition of female labor force in the market.

In a study conducted by Galor and Weil in 1996, the relationship between economic growth and birth rate is analyzed and the fertility of women and income inequality caused by fertility are emphasized. The study identifies the inequality created by the fact that high-income families have fewer children than low-income families. In the study, which is known in the literature of economy as the study defined as "birth rate-dependent income inequality" over time; it is shown that in addition to the increase in the worker's wage due to a woman in employment becoming a mother, the realization of the cost of having children more than household income results in a decrease in birth rates and an increase in the amount of capital per capita. Whether the decreased/increased birth rate increased/decreased the growth rate depends on whether quality/qualities rather than quantity/qualities determine individuals' decision to have children. In addition, it is stated that the fertility decision affects growth, depending on the level of education of the society (more precisely of the woman) (Galor and Weil, 1996: 375-377). Although Galor and Weil's study does not emphasize the exclusion of housewives from employment, it is considered to be one of the important studies showing that the situation of housewives has a dampening effect on economic growth as it increases income inequality between the sexes and, accordingly, inequality in the acquisition of property.

The study conducted by Anker et al. for the ILO in 2003 to generate data on working life does not conceptualize the invisible domestic labor of housewives, but is concerned with occupational gender segregation and the strong tendency for women and men to be concentrated in different occupations around the world. The work done by women and men is documented under the term "sex segregation", which refers to the division of paid work according to gender. In the ILO study, cultural and social attitudes towards the division of labor are differentiated as "men's work" or "women's work". By gender and occupational groups, the study, which aggregates employment data into eight professional job categories, varies from country to country and from one job to another. However, assessments focus again on employment; noting that people should be able to enter the world of paid work and choose a traditional or non-traditional occupation (and education) without having to face discrimination and other negative consequences (e.g. lower pay and thus economic dependency, sexual harassment, glass ceiling) (Anker et al. 2003: 26).

The ILO Convention No. 189 on Decent Work for Domestic Workers, published in 2011, states that *"domestic work has become devalued and invisible, and that such work is mainly done by women and girls"*. The first article of the convention defines "domestic work" as work performed for or within the home or household, while any person engaged in domestic work within a contract defining an employment relationship is defined as a "domestic worker". In this way, the ILO includes itself in the criticism that the diversification of forms of employment in the labor market and the creation of very different groups creates polarization and inequality (ILO, 2011: 3). As can be seen in this convention, although the ILO has identified women and young girls as the ones who do domestic work, when it comes to human rights, it still adheres to the condition of employment and considers women's paid work for another household as labor, whereas the work done in their own household is seen as natural housework. At the heart of the ILO's contradiction is undoubtedly the uncertainty of who will pay for the domestic work done by women and give them their humanly right and compensation for the labor they fulfil day by day. This ambiguity is also thought to represent the need for an assessment of labor independent of employment. The ILO's April 2020 report on COVID-19 and updated forecasts and analysis on the world of work states that an unprecedented contraction of employment has begun in many countries (ILO, 2020: 1). The content of the report provides a clear picture of the consequences of employment dependency in the world labor market, with changes in working hours, reflecting both "layoffs" and reductions in working hours.

It is noted that comparative analyses of women's and men's employment have been undertaken with the aim of expanding more marginal forms of employment; and that there is a need for studies that take into account women's productivity outside the area of employment; such as child-rearing and housekeeping, which are priorities for them when it comes to their own life circumstances. Women who are directed to work in another job outside the home while at the same time taking the responsibility of childcare and housework at home, are also the one's who have to manage the additional financial and moral costs of being occupied in two work areas at once. This situation, as explained in the population statistics in Turkey, is reflected in the fact that one of the reasons why the employable female population does not want to be employed outside of home, is their preference of raising children at home and being a housewife. In other words, if a housewife wants to benefit from the insurance advantage, she must have a job outside the home and leave her child and household chores to another woman/caregiver (whom she pays for as the employer). It is suggested that this sick cycle can be solved by insuring the housewife who takes care of her own child and does the housework herself in return for this labor (Gündeşlioğlu and Yıldız, 2011: 19).

In his 2020 study, Slesnick explains the difficulty of showing how any reasonable process of aggregating individual welfare functions would affect GDP as a measure of social welfare. In describing this difficulty, he notes that similar studies[18] in the literature of economy have proposed frameworks for measuring social welfare that incorporate principles of equity in national accounts. For example, in examining the evolution of inequality in income, consumption, and leisure over time, he notes that non-market factors related to household production also affect welfare in ways that are not included in GDP. Slesnick notes that the studies he examines do not take into account the potential distributional effects of aggregate production levels that would be included in the equation for social welfare functions, but they encourage a revival of interest in their inclusion (Slesnick 2020: 484-493).

In a literature analysis by Feldstein (2017), it is noted that W. Mitchell et al. (1921) proposed a "hypothetical value of housewives' services" equal to about 30 percent of more narrowly defined conventional national income estimates. Although these approaches were estimated to affect economic welfare, they were not taken into account in national income measurements; and since Kuznets' (1934, 1941) early work, national output has been defined as the exclusion of goods and services produced within the household as well as services provided outside the household but not sold (Feldstein, 2017: 4-5).

Franzis' analysis of various studies also empirically compares measures of income inequality. However, he notes that the data available for such an analysis is far from perfect and suggests that the weak relationship between money income and household production income may be due to the procedures used to overcome data deficiencies, especially the calculation of household production income, and that a stronger relationship may reverse the estimate of household production income by including household production (Franzis, 2009: 2).

Kılınç and Yetkiner's study also examines women in employment and shows that a more gender equal labor market contributes to income convergence. In the study; it is stated that policies to increase the share of women in employment should be followed (Kılınç and Yetkiner, 2013: 6-20). While Kılınç and Yetkiner's study emphasizes that gender equality in employment has a positive impact on income convergence, it is understood that changes in the children or family lives of women in employment are not taken into consideration, especially whether the same women also do household chores or the situation of children deprived of their mothers. Moreover, both of the two levels of policy outcomes for women who are defined as a "disadvantaged group" are focused only on increasing the presence of women in employment, which can be cited as an example of many studies that evaluate women's labor in terms of employment.

---

[18]For similar studies, see Nordhaus and Tobin (1972), Attanasio and Pistaferri (2016), Jorgenson and Slesnick (1987), Piketty et al. (2018).

In the United Nations' work on the status of women, it is stated that women's contribution to the well-being of nations, peoples and communities is only possible through their full participation in the areas and processes where social decisions are made. Humanity's quest for a healthy world that is just, diverse but united and offers opportunities for growth and development for all its inhabitants can only exist if women work with men to realize it. It calls for a bold shift in the broad vision and outlook of world leaders, based on the betterment of all humanity; emphasizing that lasting gender equality can only be achieved by abandoning outdated beliefs, cultural norms and practices that build on existing strengths and do not serve the common interests of humanity (UN, 2020: 3). From this point of view, the unifying power and encouraging approach of the United Nations suggests that it will take a little more time to raise awareness on the issue and for the mathematics of scientific studies to emerge, in order to achieve gender equality.

In the study assessing the sustainability of development in economic terms, Claros (2012) states that a new development model is deemed necessary. However, the mathematical expression of such a model, while far from perfect, envisages a transition to a new development paradigm that is compatible within the existing system; with features that have proven to be durable and not difficult to predict. These features include respect for property rights, progressive taxation, the use of the government budget as a key mechanism for income distribution, universal primary education and gender equality. It is also considered important to regulate and further develop the institutions that underpin an increasingly complex global economy (Claros A. L. 2012: 55-56).

Studies in Turkey (Kutlu, 2016 and Aydın, 2012) have also examined the impact of social assistance on labor supply through employment and have shown that social assistance reduces women's preference for full-time employment. While it is generally stated that being a man increases labor supply, "the opposite is the case for women"; and moreover, it is also stated that social assistance decreases full-time labor supply. In both studies, housewives' labor is not considered as labor supply.

In Kumcu's (2019) theoretical study analyzing mainstream development concepts, the main concepts and related terms of development theory are discussed to elaborate on the intersection of development, gender equality (or inequality) and labor. The study attempts to establish the paradigm that will be used as a determinant of a fair and effective development with the new concept of "labor equality". Kumcu argues that old patriarchal economic structures can be changed by accepting that the labor supply provided by housewives to their households contributes to the economy as much as regular employment, with a completely new idea for a paradigm shift (Kumcu, 2019: 45-50).

In the literature of economy, it can be observed that the place of women in the economy, who are considered to behave rationally like all individuals, is analyzed under two different headings in studies evaluating employment conditions. According to the number of studies, these topics are listed as follows.

1. Employment participation rates of the female labor force,

2. Injustice and problems women face in working life.

Especially in the literature known as women's studies, there are many studies on women's participation rate in employment and increasing it, including many public and private organizations from policy makers to the United Nations and ILO. Another issue that draws as much attention as the employment rate in studies on women is the difficulties faced by women in working life and the prevention of injustices suffered by them. However, despite the widespread use of homemaker labor, relatively little attention has been paid to the conceptual problems related to measuring the inequality it is subjected to and few contributions have been made to the theoretical foundations of the issue. The main argument of this study, that the rationality of the rational labor owner (as boundedly rational) in labor theory is altered, is almost non-existent.

In the literature of economy, there is no study that deals with the labor of housewives, who represent the limited rational population out of employment; no one considers this issue by looking at it as an independent component of employment; neither questions it nor measures its contribution to economy. Despite the fact that there are gender-related reasons underlying the labor violation of the largely unemployed female population as housewives, the multiplicity of studies focusing on increasing women's employment and the unidirectionality of the theories draw attention. It is known that women, who are at the center of intertwined problems related to both their participation in employment and raising healthy children as mothers, have many problems such as their inability to participate in governance, labor inequality and inability to defend their property rights. The root causes of these problems should be sought in women's lack of economic freedom; in other words, in the fact that although most of them are working as housewives, they cannot prove that they are working and their unregistered status continues for generations.

### 3.2. THEORETICAL EXPECTATIONS

Thomas Samuel Kuhn, in his book "The Structure of Scientific Revolutions", states that the establishment of a paradigm and thus the ability to conduct more closed and specialized research is an indicator of maturity in the development of that branch of science. However, when a synthesis is created in the theoretical schools that enable the collection of facts that give rise to science and the emergence of theories, at a level that can attract the majority of the next generation of scientists, the old schools gradually disappear. It is stated that those who adhere to the old view are gradually erased from the professional environment and no one cares about their work because a theory must be stronger than its competitors in order to be accepted as a paradigm. According to Kuhn, if scientists cannot adapt their work to the new and stricter definitions that the paradigm brings to the field of science, they will move forward alone and in the field of philosophy, which has become today's special sciences. However, in ordinary science, as the paradigm from which the research originated loses its functioning, its constraints lighten, the nature of the research questions change, and thus the scientific crisis is overcome and new theories are formed (Kuhn, 2008: 83-99).

The theoretical expectations of the study can be stated as the solution of the paradox of the GDP calculations put forward in Samuelson's book, which is taught in economics courses, regarding the results of this questioning, and going further; the expansion of a paradigm for questioning the labor-employment dependence.

### 3.2.1. Analyzing Samuelson's Paradox

This study, whose main problem is labor, uses the question of how the labor of housewives affects GDP as an important question in the inquiry. Samuelson, in the most widely taught "Economics" textbook, the first edition of which was published in 1948 and the 20th edition in 2019, starts the paradox about the level of national income under the heading "Some Special Points" at the end of the chapter "Measurement of Economic Activity". *While in the first editions of his book he expresses the fact that the goods and services produced by a housewife are never counted in national income calculations as "a master marrying his slave"*, in the 1968 edition he states: "So if a man marries his housekeeper, the national income (NNP) may collapse! Or if a wife agrees to clean her neighbor's house for $4,000 a year, NNP increases by $8,000". Samuelson states that the housewife component is not omitted for logical reasons; it is omitted because it would be difficult to obtain accurate estimates of the monetary value of a wife's services. Assuming that the number of women working at home will not change much in terms of importance, Samuelson assumes that the level of NNP will remain the same on average, with or without the inclusion of vegetables grown, meals prepared and other similar activities (Samuelson and Scott, 1968: 217-218).

In later revised editions of the book, Samuelson moves the issue from the annotations section to the upper part and presents it for discussion with the question "How does the level of GDP change when a person marries her lover or her gardener?" (Samuelson and Nordhaus, 2009: 407).

It is known that in a market economy, the aggregate demand for women's labor is accepted as an economic value when it is met on the condition of employment, while the value created in response to individual demands from households is not accepted and even ignored. Therefore, it is believed that the main reason behind the distribution debates is the uncertainty about who owns labor and what its value is.

Considering the complexity of the variables in growth theories and the convergence problems across countries, it seems that the literature on national income calculations as a common measure of economic growth has not dealt with the number of housewives working at home. Therefore, although this debate seems to have died down over time, millions of housewives who are not employed in today's market continue to produce goods and services that are not included in GDP calculations. While these products produced by housewives are not taken into account in GDP calculations, when a worker is employed at home to do the same work, the work of this worker is considered a market transaction and increases the value of GDP. The reason why the share of women's domestic labor in GDP seems to have reached a non-negligible size is overpopulation; which in turn is thought to have (re)transformed domestic production into a sector. As the expected contribution of the study to the theoretical literature, Samuelson's paradox is attempted to be answered under the heading "Results of the Model".

### 3.2.2. Evaluation of Labor as Independent of Employment

Paradigm-based research, from which the concept of labor emerges and is shaped, is thought to have led to the development of economic science while at the same time - due to the constraints arising from the reliance on the paradigm - delaying the emergence of the usual discoveries of science. The reason why this study focuses on the concept of "labor" is the scientific crisis in economics. The issue of paradigm is open to broader considerations in the field of sociology of science. However, here in the paradigm, chosen as the basis of economic theories, there is a need to question the labor-employment dependency in order to understand whether the labor of the limited rational individual in the housewife example is actually contrary to the definition of "economic value" (Kumcu, 2017: 12).

In economic terms, in order to increase its profitability, a producer must observe the difference between the output demanded by the firm and the costs it incurs. In the union established as a family, a similar approach is in question for the labor demanded by the family, such as taking care of children, cooking and other household chores performed by the woman; the mother of the house. Because in the competitive market conditions of the capitalist system, the woman, who has to make a choice between working at a job in addition to housework and fulfilling the responsibilities related to children, acts rationally in accordance with the definition of 'homo economicus' and ends up preferring to be a housewife. However, while this preference limits the woman's rationality to be the mother of the house and gives her the opportunity to perform her maternal duty in a healthier way, it does not guarantee that her professionalism or experience will be rewarded in terms of her labor. In the long run, it is observable that the upbringing of individuals who will realize and ensure the continuity of social welfare is a result of the productive labor undertaken by women within the concept of family.

While the "fertility", which has no alternative, forces women to choose to be "voluntarily unemployed" as defined by the market, labor is considered economic only when it is associated with employment, and with the interests of the actors of the capitalist economic system. As noted in the previous sections on Keynes' demand-driven economics and Veblen's consumption approaches, it is accepted that consumption is more determinant than production. Just as the focus of growth models has shifted from production to consumption, it is foreseen that labor, which is evaluated with a focus on the employment condition, can also be evaluated outside of employment while considering economic development.

Table 4 shows the percentages of labor force participation in OECD countries under employment condition - as labor-employment dependency.

*Table 5: Labor Force Participation (Employment Dependency) Rates in World Countries (%)*

| Country | Labor Force Participation | | Employment | | Unemployment | |
|---|---|---|---|---|---|---|
| Male | Female | Male | Female | Male | Female | Male |
| Turkiye | 72.7 | 34.2 | 65.7 | 29.4 | 9.5 | 13.9 |
| Italy | 59.4 | 41.1 | 83.6 | 36.3 | 9.7 | 11.8 |
| Greece | 60.0 | 44.3 | 50.8 | 33.6 | 15.4 | 24.2 |
| Mexico | 77.7 | 45.3 | 74.9 | 42.0 | 3.2 | 3.4 |
| South Africa | 62.3 | 48.4 | 46.7 | 34.4 | 25.1 | 29.0 |
| France | 60.5 | 51.8 | 55.2 | 47.2 | 8.7 | 8.7 |
| EU | 64.9 | 52.4 | 60.6 | 48.7 | 6.6 | 7.1 |
| OECD | 69.0 | 52.5 | 65.5 | 49.6 | 5.2 | 5.7 |
| Spain | 54.6 | 53.1 | 55.7 | 44.0 | 13.7 | 17.0 |
| Portugal | 64.4 | 54.5 | 60.2 | 50.4 | 6.6 | 7.4 |
| Hungary | 70.3 | 55.1 | 67.9 | 52.9 | 3.5 | 4 |
| Germany | 66.8 | 55.9 | 64.3 | 54.3 | 3.8 | 2.9 |

Source: OECD, 2019

The table is considered important as it shows that women's labor supply is evaluated according to employment conditions worldwide and that the female population appears to be employed at a lower rate than the male population. It also creates a misperception that women do not produce economic value, in other words that they consume more than they produce. However, we know that in the social culture, the prejudice that women "do not work" because they are concerned with housework and raising children is widely accepted. We believe that efforts should be made not only to change this prejudiced reality by increasing women's employment, but also to make the work of so many women truly visible; which should be the responsibility of everyone who is aware of this.

Socially and economically, the presence of the mother and the fact that she does housework is considered very important for the family. Even if the housewife, who is a permanent worker, interrupts her work, she still manages the household in general, and in this way the decision-making authority within the family can be made sustainable. Since it is out of the question for the family in the managerial position to dismiss the housewife mother, job security seems to be ensured. In this case, the positive contribution of the housewife to domestic production, whether productive or unproductive, continues. If another female worker is employed at home to do household chores, the family will have to bear more costs to ensure the continuity of this worker's work, and the condition that the return from this employment is positive in all cases will be eliminated. Moreover, if this employed worker shirks, she will be fired and her job security will be eliminated. In this scenario, it is not that women expect a monetary return for what they do for their families; rather, it is pointed out that the demand for housewife labor is decisive in the economy for the continuity of social welfare.

### 3.3. THE MODEL OF THE IMPACT OF HOUSEWIVES' LABOR ON GDP ACCOUNTS

In this dissertation, which argues for Labor-Employment Independence, the "Model of the Impact of Housewives' Labor on GDP Calculations" was developed using the Atkinson Inequality Scale. The model attempts to quantify the contribution of housewives working at home (unregistered) independently of employment to GDP and to measure the benefits they provide to continuity and employed (registered) workers and thus to economic growth.

As explained in the first part of the study, in economic theories, labor is evaluated under the assumption of full employment. In Keynesian theories, it is argued that it is possible for the state to intervene in the economy by increasing public expenditures in order to steer the economy towards full employment. The assumption of full employment also implies that all workers, whether at home or outside the home, should be paid wages. However, it is known that no such remuneration is made in practice. The capitalist appropriation of this theory has unfortunately turned into policies (maternity leave, milk leave, etc.) that have been shaped to make it possible for even mothers in the newborn period to leave their babies and work outside the home. This is the point where the capitalist understanding has almost pitted the social state and science against each other. It is known that states, even those making an effort to be a social state, are trapped by the conflict of being on the side of the employer, only provide salary/wage supports such as family allowance payments or minimum subsistence allowance, which do not meet the criteria of full employment assumed by science, and which is commonly only payed to the registered working parent.

In a society dominated by capitalist relations of production and property, if it is not possible for productive labor in the capitalist mode of production to reproduce itself in a widespread manner and without being subject to productive labor, it is also not possible for productive labor to reproduce itself in a similar manner and without being subject to non-productive labor. For example, in a family with two children, where the mother is a housewife and the father is a construction worker, the father's continuity in work is not possible if the mother's housework is not done regularly and the children are not kept safe. Likewise, if the father does not earn money by working at a job, it is not possible for the mother to ensure continuity as a housewife in meeting the needs that cannot be produced at home. Considering that what is vital is to ensure the continuity of the human generation, not the continuity of capitalism; the man supports the family's livelihood by being employed in a job to meet the needs that can only be acquired in exchange for money, apart from what his wife, the housewife, produces.

The issue of economic growth, as is well known, forms the basis of economic development (development) and has an important place in measuring the development of countries. When the problems related to economic development are analyzed, it can be seen that it is capital-oriented but still constantly demanding regulatory support from the state; and moreover it is market-oriented and dependant on gender differentiation. Therefore, the main issue discussed in this study is how it is possible to take into account the labor of housewives as limited rational individuals in order to clarify the meaning of the concept of labor, due to their role in ensuring the continuity of human capital, which increases the labor supply in the long run.

As stated by Todaro and Smith (2012), achieving the goals set as development indicators depends on the increase in the number of individuals who feel valued, have increased self-esteem and are self-aware. These individuals are defined as women and men who know what they want and can make choices among the freedoms and different alternatives offered to society (Todaro and Smith, 2012: 206).

Women and girls are thought to be unable to contribute fully to society, either as a natural consequence of being prevented from owning the land they work for or because social norms make them dependent on male relatives to participate in the economy. But the path from doubt to confidence, from silence to voice, from passivity to action should not be understood as simply entering the labor market or fitting into a global production chain of one kind or another. Capacity development must deal with all aspects of human existence - social, intellectual, cultural, spiritual, moral as well as economic. It is recognized that such questions are as critical to the human implications of scientific development as they are to women's empowerment efforts. Decades of experience have shown that when increasing numbers of women and men, young and old, from all economic and educational backgrounds, work together to learn about the patterns of relationships and corresponding social structures that reflect the fundamental unity of the human family, then only then it is possible to bring about change that improves society. Drawing on the experience of many, it is envisioned that creating spaces and mechanisms for individuals to share science and life will open up lasting avenues for universal participation and indispensable pathways for processes of social change (BIC, 2018: 3).

In the strategies of the United Nations Development Program, which continues its practices worldwide, the economic development of countries is linked to the principle of equality among individuals. It is stated that achieving sustainable development and eliminating poverty is possible through the efficient use of resources and the empowerment and inclusion of women in the economy (UNDP, 2008: 12).

As explained in this section of the study so far, although there are different components in each of the GDP calculation methods, it is noteworthy that the common element is people. It is suggested that the portion representing the labor of housewives, which is not shown in these calculations, should be taken into account as the equivalent of all goods and services consumed by households.

### 3.3.1. Model Introduction and Methodology

In this study, the method to be used in measuring the economic return of housewives' labor consists of a two-stage model. The Atkinson Inequality Scale, a normative scale that measures inequality aversion and sensitivity to inequality, is used because housewives, who provide unpaid labor as limited rational individuals, do not have any income and families show different rankings depending on inequality measures such as income level, age and number of children. A sample calculation is presented for only one selected year with the Atkinson Scale, which makes it possible to measure inequality using raw data.

In the first stage of the model, the Atkinson inequality measure, I1, is calculated for the current situation using the GDP calculated by the income method for the selected year. In the second stage of the model, the Atkinson inequality measure, I2, is calculated for the new GDP calculated by including housewives' labor and our hypothesis that housewives' labor is an economic quantity will be tested by comparing the values of both cases.

As can be understood, the model is designed to show that the consumed labor of limited rational individuals in a country (in this study, housewives in Turkey and in the GDP calculated for the past year using the income method) should be taken into account rather than recalculating the GDP.

The originality of the model is that the calculation using the Atkinson Inequality Scale shows that the economic input consumed as labor or labor force should be replaced by spending a certain income which is human energy. In the model, it is shown by using the Atkinson Inequality Scale that the human energy that is consumed as an economic input should be replaced by spending a certain income, which is the labor of housewives, considered to be included in the social welfare function.

It is believed that this model and calculations will lead to other new studies with different theoretical expansions required by the paradigm, considering that labor is a consumed energy and the limited rationality of its owner.

### 3.3.2. Introduction of the Atkinson Inequality Scale Used in The Model

The Atkinson Inequality Scale (hereinafter referred to as the Atkinson Scale), as a normative inequality scale, is based on the idea that a decrease in the welfare of the society depends on an increase in income inequality and is used to measure the loss in social welfare caused by an increase in income inequality. The scale was first proposed by H. Dalton[19] in 1920. Later, it was developed by Anthony B. Atkinson in 1970 in his article "On the Measurement of Inequality" by adding the sensitivity of society to income inequality to the welfare function. It is known as the most widely used normative criterion for measuring inequality (Kaya, 2018: 25).

---

[19]Dalton's approach in *The measurement of the inequality of incomes* was applied in Great Britain in 1919-1920 and equal distribution of income was achieved.

Atkinson takes a new approach to inequality measures that economists use when focusing on conceptual problems related to income distribution and welfare to answer a wide range of questions. Stating that traditional approaches in almost all empirical studies adopt statistics such as variance, coefficient of variation or Gini coefficient, Atkinson argues that the basic problem underlying such calculations is some kind of social welfare concept and that this concept should be approached by directly addressing the form of the social welfare function to be used (Atkinson, 1970: 244).

Atkinson obtains a measure of inequality by comparing two frequency (fy) distributions of y with the function f(y) representing the frequency of the income distribution of a characteristic 'y', which he calls income. Two different objectives can be achieved by comparing the distributions. First, a simple distribution ranking can be obtained, such as determining whether after-tax income is more evenly distributed than pre-tax income; secondly, going further than the first, it is possible to measure the difference in inequality between the two distributions (Atkinson, 1970: 245).

Referring to Dalton, Atkinson defines the welfare (W) function as the sum of the utility functions of all individuals in society with decreasing marginal utility with respect to income and ranks the distribution according to equation (1);

$$W = \frac{1}{N} \sum_i^n U(yi)$$

$$W = \int_0^{\bar{y}} U(y)f(y)dy \qquad (1)$$

In order to arrive at any distribution ranking, Atkinson makes some assumptions about the form of the function U(y) and ranks two distributions, bounded primarily by the increasing and concave function, as acceptable requirements, without further specifying the form of the function U(y). At this stage, Atkinson, drawing on different studies on the problem of decision making under uncertainty, states that the ordering of income distributions according to equation (1) is statistically identical to the ordering of probability distributions f(y) with respect to expected utility and that the assumption that U(y) is concave, is equivalent to assuming that a person is risk averse. Atkinson follows Dalton in assuming that the utility function U(U(y)) is twice continuously differentiable and that it will be an additively separable and symmetric function of individual incomes. Thus, for U(y), (U' > 0, U'' ≤ 0), a distribution f(y) will be preferred to another distribution f*(y) according to equality condition (1), where F (y) = $\int_0^y f(x)$ dx;

$$F(y) \neq F^*(y) \qquad (2)$$

Here, condition (2) refers to the diminishing marginal returns to income of the utility function, which is known as a strictly concave function in the literature of economy. The paper argues that restricting incomes to the range $0 \leq y \leq \bar{y}$ is not too restrictive and mathematically appropriate when it comes to the welfare problem. Here, denoting the mean of the distribution as (μ) (Atkinson, 1970: 245-249);

$$W = \int_0^{\bar{y}} \frac{U(y)f(y)d(y)}{U(\mu)}$$

$U(y)$ denotes the utility obtained from his income, and $U(\mu)$ denotes the total utility in equality.

Assuming that the distributions are compared with the same mean, Atkinson obtains a measure of inequality that does not change according to linear transformations by introducing the concept of a level of per capita income equivalent to an evenly distributed income level yEDE. If the current distribution is equally distributed, it will yield the same level of social welfare (Atkinson, 1970: 250). In other words;

$$U(y_{EDE}) \int_0^{\mu} f(y)dy = \int_0^{\mu} U(y)f(y)dy$$

In this case $I$ can be defined as the new inequality measure;

$$I = 1 - \frac{yEDE}{\mu} \quad (3)$$

$I$ is equal to the difference from 1 of the ratio of the level of income equivalent to the equally distributed income to the average of the actual distribution. Income redistribution can be planned so that we can say that equally distributed income would increase social welfare by an amount equivalent to a 5% increase. This makes it easier to compare the gains from redistribution with the costs - such as the disincentive effect of income taxation - and the benefits of alternative economic measures. Noting that the notion of income equivalent to equally distributed income is closely related to a risk premium or certainty equivalent in the theory of decision-making under uncertainty, Atkinson recalls that almost all measures traditionally used are concerned with measuring inequality independently of the average income level. In this case, when the income distribution in country A is regarded as simply a magnified version of the income distribution in country B, $f_A(y) = f_B(\theta y)$, characterized by the same degree of inequality, the equally distributed measure $I$ must be invariant with respect to these proportional shifts in order for the degree of inequality to be considered independent of the average income level (Atkinson, 1970: 251).

Atkinson differs from Dalton at this stage by showing that inequality depends on the choice of ε; a measure of the degree of inequality aversion or sensitivity to different income levels. As ε increases, more weight is given to transfers at the lower end of the distribution and less to transfers at the upper end.

At one extreme, the limiting case is defined as ε → ∞, which gives a $\min_i\{y_i\}$ function that only takes into account transfers to the lowest income group (and is therefore not strictly concave); at the other extreme, ε=0, which gives a linear utility function that ranks distributions only by total income. According to Atkinson, the sum of the utility functions of individuals constitutes the welfare of society. Considering that the welfare of society depends on ε, which he defines as the degree to which society avoids inequality; Atkinson shows this sensitivity in equation (4). The society's insensitivity to income inequality takes the value $0 \leq \varepsilon \leq \infty$ and as it moves away from zero, it is understood that the society's sensitivity to inequality increases; it avoids inequality (Atkinson, 1970: 257).

In Equation (4), it calculates a new measure of inequality, I, which is a new measure of inequality indicating the level of social welfare, averaged over the true distribution (μ):

$$I = 1 - \left[ \Sigma_i \left( \frac{y_i}{\mu} \right)^{1-\varepsilon} f(y_i) \right]^{1/(1-\varepsilon)} \qquad (4)$$

The I measure ranges between 0 (perfect equality) and 1 (perfect inequality) and is highly sensitive.

A lower value of I indicates that the distribution has become more equal, so that a higher level of equally distributed income (relative to the average) is required to achieve the same level of social welfare as the actual distribution. For example, I = 0.30 allows us to say that if incomes were evenly distributed, we would only need 70% of current national income to achieve the same level of social welfare (according to the particular social welfare function) (Atkinson, 1970: 250).

The Atkinson Scale is used in many fields related to the measurement of inequalities. Especially in the field of economics, it is used to measure inequalities in income and wealth distribution between countries or regions and the welfare levels that occur as a result of these inequalities.

In the analysis of national and international literature on the use of the Atkinson Scale by Çiftçi (2010 and 2018), it is shown that there are studies on measuring inequality in regional income distribution and measuring regional differences in regional productivity, as well as wages, personal income, and regional differences in financial products, loans, capital stock and agricultural yields (Çiftçi, 2018: 406 and Çiftçi 2010: 1390).

The Atkinson Scale is also interpreted as a measure of the potential gains in redistribution in terms of showing how much difference will be given up when switching from income distribution practices in society to absolute equal distribution (Öztürk and Göktolga, 2010: 4).

In the Atkinson inequality measure, which is directly derived from welfare theory, individual incomes are included in the social welfare function with diminishing marginal utility. Equally distributed equivalent income then represents the level of per capita income that, if shared equally, would produce the same level of social welfare as the observed distribution. Based on this concept, the Atkinson inequality measure is defined as the percentage of average income lost due to inequality (with respect to equally distributed equivalent income).

The exact form of the Atkinson inequality measure depends on the parameter ε of inequality aversion in the social welfare function. Values of 0.5, 1 and 2 are commonly used for ε, as in studies for the United Nations. It is argued that the Atkinson measure is more advantageous than the Gini coefficient because it can decompose inequality within and between groups (Harttgen and Volmer, 2011: 8).

### 3.3.3. Limitations and Assumptions of the Model

This study focuses on the labor of housewives, who are unpaid labor producers, through its impact on GDP in order to demonstrate that labor is an economic value irrespective of employment. The broad challenges with labor markets, the linked issues with women employees in the workplace, and problems with employment legislation are thus left out of the study's scope.

Women who work as housewives are thought to adapt to one of five situations: those who are unemployed, those who work part-time, those who prefer not to work in another job because of their domestic responsibilities, those who work outside the home as unpaid family workers, and those who have a full-time job (employed) but also take care of the household. However, in order to remove the suspicion of double counting in the income accounts, we restrict the study to the female population of housewives, defined as limitedly rational and "not in employment solely because of housework".

It is well known that it is difficult to show how household production and other possible related factors will affect GDP, which is considered as a benchmark for measuring social welfare. Therefore, the model used in this study has assumptions that determine the inclusion of variables in the hypotheses for social welfare accounts.

Although in need of discussion and improvement, the assumptions of the model, which are considered important for adapting macroeconomic data to individual labor and income, can be listed as follows:

1. Rapid population growth has shifted the fulfillment of individual needs from the micro to the macro scale, thus turning housework into a sector.

2. Except for families consisting of upper-income individuals, housework is performed by the housewife, who is the mother of the household.

3. All housewives are potentially able to do all household chores.

4. All goods and services produced by housewives at home are consumed.

5. What is consumed is the labor of the housewife, and there is a minimum wage paid for the labor.

6. Household chores, including baby/child care, are the same as those performed by those employed as domestic workers in the market.

7. The distribution of national income and household rankings of housewives are logarithmically and normally distributed.

Since the issue of the minimum wage is related to labor market analysis, the amount of the minimum wage, who pays it, who does not pay it, or its informality are not questioned within the limits of this study. Men who are married or unmarried but work for their families are not included in the study as they are considered to be exceptions, as stated in TURKSTAT data.

### 3.3.4. Model Variables and Data Set

As explained in the literature of economy, focusing on increasing women's employment for economic growth, all housewives, simply by virtue of being women, are expected to work, albeit at low wages, and also to take care of their children and husbands. However, under these conditions that limit their rationality, housewives prefer to employ themselves in their own homes rather than in a job. In an economy where all workers demand the same rights or where no woman does housework as a housewife, it will not be easy or possible for the employer to find a replacement for a worker who is disruptive.

The above considerations bring us to the stage of determining the labor paradigm and the variables to be used in the calculations of this study. These variables are listed below, along with the data used in the model and their symbols. Housewife labor, denoted by $L_{Fhw}$, is defined as the independent variable of the study and shows the size of the female labor force working informally at home as a selected limited rational individual. NKev denotes the population of housewives used in the calculation of LKev.

As explained in the assumptions section, NKev consists only of "*women who do not work in another job due to housework*" and is used as given in Table 6; which shows the reasons why individuals are not in the labor force. Women who are not in the labor force or who do housework on a part-time basis; Women who do not work in another job due to housework; women who work outside the home as unpaid family workers... The labor of women who cannot find a job but do housework themselves is not taken into account to avoid double counting, due to the possibility that they may have earned income.

In the calculations of the model used in this thesis, household and employment data and macro data on GDP and income distribution prepared by TurkStat for Turkey are used. Since the series for housewives is not comparable with previous years due to the new arrangements made in TURKSTAT statistics since 2014, data for 2014 and later years is available. Although it is possible to calculate the effect of housewives' labor on GDP for any year in the model, a sample calculation is made here by selecting the year 2014 since the statistics regulated since 2014 are accessible. Housewife labor, which is the main variable of the model and denoted by LKev, is defined as the magnitude of the effect of housework on GDP. In the first stage of the model, LKev is calculated in terms of income using NKev and the gross minimum wage ($P_A$) for the selected year, as shown in Equation (5):

$$L_{Kev} = N_{Kev} \text{ x } P_A \hspace{2cm} (5)$$

Table 5 shows the average income μ calculated for the calculation of the Atkinson inequality measure I1, which is the inequality in the current situation in the first stage of the model, and TURKSTAT data for 2014. Average income μ is calculated by TurkStat as per capita income in the national income accounts of the relevant year; however, it can also be obtained by dividing GDP by the number of people. Income and minimum wage information obtained from TurkStat data are shown in Table 5, organized as follows:

**Tablo 6:Data Used to Calculate Inequality Measure I1 for the Current Situation**

| Population, Income and Total Product Information | | 2014 |
|---|---|---|
| Total (men and women) Population in Employment(Thousands) | $N_L$ | 25,933 |
| Total Population (Thousand people) | $N$ | 77,695 |
| Gross Minimum Wage (TL/Month) | $P_A$ | 1,102 |
| GDP per capita at current prices (TL/Month) | $\mu$ | 2,204 |
| Output Excluding Housewife Labor (Thousand TL) | GDP | 2,054,897,828 |

Source: Created using TurkStat 2020 data

Total population in employment (men and women) NL is taken from Table 7, Total Full and Part-Time Employment; total population N is taken from Table 9, *Population by Gender and Age Group;* and gross minimum wage PA is taken from Table 10. GDP for 2014, calculated by TURKSTAT using the income method, excluding housewives' labor, is used as given in Table 11. Tables of TurkStat data can be found in the Annex at the end of the study.

The total number of male and female wage earners in employment is denoted by NL. GDP, which is the sum of the incomes of these workers, is defined as the dependent variable of the study. In its current form, GDP excluding housewives' labor is denoted as LGSDP and used in the calculation of I2 to indicate the addition of LKev in the measurement of inequality.

$$L_{GSYH} = GSYH + L_{Kev} \qquad (6)$$

Since this study aims to measure the inequality caused by the unpaid work groups of the population and their contribution to inequality, it is important to pay attention to the integrity of the components of total income that will affect the average. Atkinson makes inequality measurable by taking the same average as a reference. Therefore, $\mu_{Kev}$, which is defined as the average income, should be used in the following inequality calculations, i.e. those that take into account the labor of housewives. In the model, when calculating $I_2$, the average income of all labor contributing to the generation of $L_{GDP}$ is considered as $\mu_{Kev}$. The calculation method is the same as for $\mu$, which is the average per capita income of the GDP distribution. μKev is obtained by dividing the total GDP by the total population, $L_{GDP}$, which is the total product of the inclusion of housewives' labor, $L_{Kev}$, in GDP, as shown in Equation (7):

$$\mu_{Kev} = \frac{LGDP}{N} \qquad\qquad (7)$$

The tables of TurkStat statistics from which the components of the labor supply situation as a housewife are derived, can be found in the Appendices section at the end of the study. Due to the fact that these components were obtained from different tables, the relevant data were checked by cross-counting to avoid duplication; it was verified that they were internally consistent and that there were no repetitions. As can be seen in the footnotes of the tables containing the raw data, there are minor differences due to rounding of some data by the data provider.

### 3.3.5. Hypotheses of the Model

In order to measure the contribution of housewives' labor to social welfare, its effect on GDP is questioned. A product that is in demand to meet a need in economic life is valuable and the labor spent on it is also considered valuable. Inequality measures will be calculated and compared for the current situation where housewives' labor is not taken into account in national income calculations and for the proposed situation where it is taken into account. With this comparison, the effort will be made to show that while the goods and services produced by housewives at home are consumed by family members, they also contribute to social welfare.

First, the inequality measure of GDP will be calculated as $I_1$ for the status quo where the labor of housewives is not taken into account in national income calculations. Then, in the second stage of the model, the inequality measure $I_2$ of GDP + $L_{Kev}$ = $L_{GDP}$ will be calculated as the inequality measure of GDP+ LKev = $L_{GDP}$ for the new total revenue that would be generated if housewives' labor, $L_{Kev}$, is taken into account in the national income accounts. Hypotheses will be tested by comparing the calculated I1 and I2.

Hypothesis H0; $I_1 = I_2$

GDP = $L_{GDP}$ Housewife labor does not affect GDP.

If there is no change in the Atkinson inequality coefficient I when housewives' labor as female labor force is included in the labor force involved in the formation of total GDP of $L_{Kev}$, H0 will be accepted by saying that housewives' labor does not affect total output.

Hypothesis H1; $I_1 \neq I_2$

GDP ≠ L$_{GDP}$ Housewife labor affects GDP.

If there is a change in the Atkinson inequality coefficient when housewives' labor as female labor force is included in the labor force involved in the formation of total L$_{GDP}$ of L$_{Kev}$, H1 will be accepted by saying that housewives' labor affects total output.

## 3.4. APPLICATION OF THE MODEL

Within the scope of the study, in order to determine the inequality arising from the exclusion of housewives' labor in GDP, it is calculated with 2014 data. In the first stage, the inequality arising from the current situation in which housewives' labor is not taken into account (I1) will ~~be~~ tried to be measured; in the second stage, the inequality (I2) that occurs if housewives' labor is taken into account will ~~be~~ tried to be measured. The hypotheses will be tested by comparing the results obtained.

This study, which proposes a humanist model to question whether housewives' labor is an economic asset or not, can also be applied to time series for more comprehensive analyses. In this preliminary study, the calculations made for 2014 are presented with comparative analyses in the findings section.

### 3.4.1.  First Stage: Calculation of Inequality Measure I1

In the first stage of the model, the Atkinson inequality coefficient I$_1$ is calculated using equation (4) in order to determine the inequality arising from the exclusion of housewives' labor in GDP for 2014. The εused in the calculations indicates the sensitivity of the society to inequality. The value of ε, which is widely used in the literature, is taken as 2 in the calculations in this study.

$$I = 1 - [\Sigma_i (\tfrac{y_i}{\mu})^{1-\varepsilon} f(y_i)]^{1/(1-\varepsilon)} \qquad (4)$$

As the value of $y_i$, the value of I1 will be calculated by dividing the minimum wage P$_A$ in 2014 by the average income μ of the distribution as the value of GDP per capita derived from GDP.

The value of $I_1$ is calculated as follows:

$$I_1 = 1 - [\Sigma_{2014} (\tfrac{PA}{\mu})^{1-2}]^{1/(1-2)}$$

$$I_1 = 1 - [(\tfrac{1.102}{2.204})^{1-2}]^{1/(1-2)}$$

$$I_1 = 0{,}50$$

It is understood that the level of social welfare in 2014, when the labor of housewives is not included in the GDP, the minimum wage PA is 1,102 TL, and the GDP per capita is 2,204 TL, can be achieved with only 50% of the national income in Turkey. In other words, an I1 value of 0.5 indicates that the inequality in income distribution is 50% in the current situation where the labor of housewives is ignored.

### 3.4.2. Second Stage Calculation of Inequality Measure $I_2$

In the second stage of the model, the social impact caused by unpaid labor supply is calculated. For this purpose, $I_2$ will be calculated using the new average $\mu_{Kev}$ calculated by taking into account the $L_{Kev}$ of housewives' labor and the hypotheses will be tested by comparing it with the value of $I_1$.

After finding $I_1$ to be 0.50, the calculation of $I_2$ requires first the calculation of $L_{Kev}$ of housewives' labor. $L_{Kev}$ is important as it will be used to calculate the new mean $\mu Kev$ and the inequality measure $I_2$. $L_{Kev}$ is calculated using Equation (5) as follows:

$$L_{Kev} = N_{Kev} \times P_A$$

$$L_{Kev} = 11.589.000 \text{ kişi} \times 1.102 \text{ TL}$$

$$L_{Kev} = 12.771.078.000 \text{ TL}$$

The information needed to calculate $L_{Kev}$ and $I_2$ is shown in Table 6. $\mu Kev$ is calculated according to Equation (7), first annually and then monthly, as follows:

$$\mu_{Kev} = \frac{LKev + LGSYH}{N} = \frac{(NKev \times PA) + GSYH2014}{N}$$

According to annual data;

$$\mu_{Kev} = \frac{(12.771.078.000) + 2.054.897.828.000}{77.695.900} = 26.612 \text{ TL/Year}$$

Monthly $\mu Kev = 26.617/12$

$\mu Kev = 2,218$ TL/month.

In the model, $\mu Kev$ is considered as the average income representing the distribution of GDP across all labor power in a country.

$$I_2 = 1 - [\Sigma_{2014} (\frac{PA}{\mu Kev})^{1-2}]^{1/(1-2)}$$

$$I_2 = 1 - [(\frac{1.102}{2.218})^{1-2}]^{1/(1-2)}$$

$$I_2 = 0,49$$

The fact that $I_2$ is smaller than $I_1$ indicates that the distribution will move away from inequality if housewives' labor is taken into account.

A distribution with an $I_2$ of 0.07 indicates that inequality can be reduced by taking into account the labor of housewives to achieve the social welfare level. The inclusion of housewives' labor in the calculations means that 51% of national income would be distributed. It shows that in 2014, when the minimum wage PA was 1,102 TL, housewives' domestic labor created a level of social welfare equivalent to 2,218 TL of GDP per capita. From these calculations, it is understood that with an increase of 14 TL in the average income per capita from GDP in 2014, the increase in social welfare reflected on the entire population by domestic work performed using housewives' labor was 1%. In 2014, when the unit price of a loaf of bread and a liter of milk was 1 TL, the Dollar/Turkish Lira exchange rate was 2 TL, the price of a kilo of meat was 25 TL, and the inflation rate was 8% and the net minimum wage was 787 TL; an income difference of 14 TL per person per month (1.8%) is significant.

This rate, which shows that the inequality in national income and distribution calculations has decreased from 50% to 0.49%, indicates that an increase in welfare is achieved only through the labor of women who are not in employment due to domestic work, and that this corresponds to the distribution of 51% of the national income in Turkey.

The study can be expanded by taking into account the labor of women who do not work only because of housework, but also those who do not find a place in employment or work part-time but supply labor through domestic production and who identify themselves as "I do not work, I am a housewife". Here, we are not talking about the commodification of labor, but about the meaning it has in human life, not only economically.

Even if housewives are not currently paid an amount in return for the labor they supply in response to the family's demand for labor, it is understood that they contribute to the real level of welfare experienced. It seems that it will take time for the mentality to be awaken of how to pay or who will pay for it, so that this conception can be developed. What we want to emphasize here, is not the presence or absence of policies such as social assistance and family subsistence allowance implemented by political governments, but the scientific definition of the concept of labor within the paradigm shift that is a social reflection of population growth.

### 3.4.3. Results and Testing Hypotheses

In the application part of the study, which includes the sample calculation of the model, firstly, the inequality measure of GDP, which is the current situation in which housewives' labor is not taken into account in national income calculations, was calculated as I1 and found to be 0.50.

The inequality measure of the new total output, GDP + $L_{Kev}$ = $L_{GDP}$, is calculated as $I_2$ and found to be 0.49 by adding $L_{Kev}$, which is calculated in the second stage, to GDP.

When housewives' labor as female labor force is included in the labor force involved in the formation of the total GDP of LKev, there is a decrease in the Atkinson inequality coefficient I, which implies a shift away from inequality. In this case, it shows that housewives' labor affects total output.

Comparing $I_1$ with $I_2$;

Since $I_1 = I_2$ (0.50 ≠ 0.49), Hypothesis H0 is rejected.

Therefore, Hypothesis H1, which asserts the existence of housewives' labor in a country's GDP and satisfies the condition $I_1 \neq I_2$, is accepted.

The fact that the $L_{GDP}$ with the inclusion of housewives' labor is larger than the GDP without the inclusion of housewives' labor, indicates a more equal distribution. It is shown that the presence of housewives' labor in this distribution already generates a welfare $I_2$ of 0.49.

Atkinson (1970) examines the problem of measuring inequality in the distribution of income (alternatively consumption or wealth). Stating that this problem is usually approached using summary statistics such as the Gini coefficient, variance or relative mean deviation, Atkinson claims that the traditional approach is misleading. According to Atkinson (1970), first, these measures often serve to obscure the fact that one cannot arrive at a complete distributional ranking without specifying the exact form of the social welfare function. Secondly, an examination of the social welfare functions implicit in these measures shows that in some cases they have characteristics that are unlikely to be acceptable and that, in general, there is no reason to believe that they will be compatible with social values. For these reasons, Atkinson hopes that these traditional measures will be rejected in favor of direct consideration of the characteristics we want the social welfare function to exhibit (Atkinson, 1970: 262).

Atkinson's approach of taking social welfare into account in order to eliminate economic inequalities seems to be in line with the early 20th century findings of Keynes, who saw the need for state intervention in the economic functioning in favor of the capitalist rather than the public. As a matter of fact, the Atkinson Scale also confirms the inconsistency and contradiction of the traditional measures of national output, which have been used since Kuznets' (1934, 1941) first study to exclude domestically produced goods and services from the measurement of national income, although they are predicted to affect economic welfare.

## 3.5. THEORETICAL AND EMPIRICAL FINDINGS

The findings of the theoretical and empirical analysis that labor can be evaluated independently of employment are listed below. The theoretical findings are based on the development of heterodox theories and the questioning of the rationality of the individual in the face of the inadequacy of orthodox economic theories that require the questioning of economic welfare. Empirical findings are derived from calculations showing the contribution of housewives' labor to GDP. The theoretical and empirical findings of the study are as follows:

- Expanding economic activities diversify the market phenomenon. Parallel to this diversification, increases in the labor demand and labor supply required to produce goods and services, inevitably strengthen and increase the labor-employment relationship. So much so that labor, as a commodity supplied by households in exchange for wages and demanded by firms to be paid for, has become the most important component of the capitalist market.

- The understanding that emphasizes the issue of distribution as the fundamental problem of economics is a consequence of the neoclassical approach, which includes the need to protect property rights. Acquiring more property means having more rights in the market. In such a class society, housewives who work for their families cannot own property.

- Going one step further from the commodification of labor and making money more important than labor starts with the Monetarist theory. Policies implemented to reduce the natural unemployment rate and increase the working population have been effective in directing housewives to employment and have turned into efforts to increase women's employment all over the world.

- The reasons why women's rights are on the agenda in the world, begin with the liberation of women from domestic slavery and their inclusion in the "working class" in employment. The idea that if a slave woman marries the master of the house, her labor will be deducted from the national income accounts can only be a valid argument for the slavery period. Samuelson's contemporary Marx, despite his opposition to slavery, also devalued women's labor and domestic work by including them in the working class according to the decisions of the owner of capital and focused only on the condition of employment.

- In the early 20th century, the rapid increase in the world's population, which reached two billion in the early 20th century, led to crises caused by the accumulation of capital due to excessive profitability caused by the transition to mass production technologies, which led to the questioning of the relationship between labor and wages. It can be observed that labor, which started to be organized by trade unions, has become a mass producing and consuming force. Population growth has also placed the human being at the center of scientific philosophy and has given rise to new economic approaches that seek to take the authority to use labor away from the capitalist and return it to the laborer. Behavioral economics, being one of these approaches, explains that human beings, who have to adapt themselves to changing conditions in order to survive, turn into limited rational individuals. In this study, a woman who finds silence as a solution to the ongoing exploitation of labor in the patriarchal economic system, is identified as a bounded rational individual.

- With the philosophy of science focusing on human beings, the scientific paradigm has also begun to stir. The responsibility of the philosophy of economics to meet human needs is integrated with its impartiality in the development of appropriate policies to ensure social welfare. Rather than evaluating the supply and demand of labor in economic functioning by considering only the interests of the owner of capital, there is a need for approaches that are in line with the impartiality of science. Especially in economic models that try to explain growth, it can be observed that the labor-employment-growth relationship has not yet been fully explained by considering the interests of all parties. When we move from income calculations to distribution calculations in growth models, the fact that housewives' labor is stated as a non-employment labor factor and is not allocated an allowance, despite the fact that it is implicitly included in the household, suggests that all calculations on distribution are incomplete or erroneous.

- When considering human development in human societies reaching billions, some conceptual gaps appear in the economic system when we move away from the division of labor according to gender and think in a human-oriented manner. It has been observed that one of these concepts that creates a gap in economic welfare is the labor of housewives. The labor of the housewife, who is the carrier of the vitality that ensures the continuity of the generation and is the provider of the basic needs of her baby, is determined as the basic labor force that ensures this continuity. In this case, the labor of a housewife is the labor of an individual who prefers to work at home to provide for her family. Based on the human-oriented approach of economics, "housewife labor" is defined as the labor spent by women in fulfilling their domestic responsibilities, similar to the labor of a worker who fulfills the responsibilities of hiring labor power to earn a living.

- It has been determined that in order for economics to fulfill its responsibility to ensure welfare and to contribute to global welfare efforts, it is necessary to measure the quantities of production and distribution accurately and that it is important to take into account all inputs in the calculation of GDP, which is used to measure these quantities. It is understood that one of the reasons for distribution problems is that housewives' labor, which is unpaid labor, is not included as labor input in GDP calculations.

- In the literature review on the subject, there is no study that deals with the labor of housewives, who represent a limited rational population by staying out of employment; and no one considers this issue by looking at it as an independent component of employment; neither questions it nor measures its contribution to economy. It is clear that the root cause of many problems such as management, labor inequality and inability to defend property rights is the lack of economic freedom of women. In other words, it has been determined that although a large proportion of women work as housewives, the fact that they cannot prove that they work, creates economic inequality.

- In order to understand the reasons for the lack of economic return, as for the labor of limited rational individuals in the housewife example, developments in the historical process were analyzed. In the scientific paradigm in which the concept of labor was born and shaped, it was determined that labor was considered to be dependent on employment.

- It has been questioned whether the labor of housewives, who provide unpaid labor as a limited rational individual, is an economic quantity and a modeling has been made. In the model, calculations made with the Atkinson Scale, which makes it possible to measure inequality using raw data, have shown that housewives' labor, which is thought to be included in the social welfare function, is an economic input that is consumed as human energy and must be replaced by spending a certain income.

Since classical economic theory, the assumption of "full employment" on which the labor-employment interdependence is taken as a base, can only be valid if the individual and the social benefit provided by the labor of housewives working at home is taken into account. Otherwise, there is a surplus of production in the market that reaches the employer through the worker as a cost-reducing benefit and is consumed. This surplus is composed of goods and services produced by housewives at home. Productive labor, which is explained above as labor that creates capital accumulation, is in fact the labor of the housewife who provides for her husband's needs at home so that he can go to work without interrupting his work.

- The consistency of focusing on human needs in terms of economic philosophy forces us to take into account the high rate of population growth over the last two centuries. If the world population were still below one billion (recall Malthus's strict population policies), capital accumulation would not have increased so much, the problem of distribution would not have grown up to these levels and of course, the debate about whether housewives contribute to economic growth would not be so important from a macro perspective, as Samuelson thought. In the global world society, the continuity of the labor supply and capital accumulation of the workers employed by an employer is thought to be fed by the continuity of the unpaid labor provided by housewives.

In human civilization, which continues its social evolution, it may be possible to achieve economic prosperity through the strengthening of civil society by ensuring that women receive their share of their contribution to the whole, and through a compromise in which both the state and the owner of capital will act together within the integrity of the individual-society-institution.

## CONCLUSION and EXPECTATIONS

In the theoretical analysis in the first part of the study, the evolutionary development of the concept of labor is examined and the results of this analysis are given at the end of the chapter. Economics became an independent discipline only in the 18th century, after the basic factors that constitute it became clear within the Scholastic doctrine. One of the searches in the process of meeting human needs has been to determine the value or price of products. The beginnings of these evaluations, which continue today, found their first answers in classical theory.

Classical economics is the period when the rationality of economic actors - most notably households and firms - was discovered. Rationality is characterized as an understanding of self-interest that claims that individuals act in their own interests, taking into account what they benefit or do not benefit from, and what they suffer or pay for. Firms seek to maximize their profits and individuals seek to maximize their benefits. Therefore, there is a natural order and equilibrium due to the complementary role of producers and consumers who fulfill these mutual needs from each other. In other words, in this economic functioning, each participating actor can meet it's basic economic needs without requiring the state to intervene. However, it was not foreseen that, the "Laissez Faire" system will accentuate injustice, instead of creating balance, as the number of human communities continued to increase. It is obvious that this was not the priority of interest groups. It was also unforeseeable that in the chain of actors from production to consumption, lawmakers, decision-makers, intermediaries, and capital owners would gradually become mostly men; shaped in favor of differences, and turn into a social problem that has persisted to the present day.

According to Smith (2008), in a capitalist economy, there is commodity production, i.e. production for the purpose of exchange (trade), the accumulation of capital in certain hands and the ownership of land. Such a society consists of three different interest groups: capitalists who own capital, workers who own labor, and landowners. Smith does not make any statement about the consumption of the unsold surplus of the product produced by the laborer within the family; the family members are the ones who consume this surplus to ensure their survival without paying any price. When it is considered that this production is produced to meet the needs of the family and not as a surplus produced for sale and not sold, it is understood how women's labor, which is not included in the capitalist perspective, is ignored in this way. In other words, although the economy is indeed at full employment, some labor is not paid for, because the laborer is a woman. It is labor that produces capital and therefore Ricardo, who defined capital as "indirect labor", made the discovery that human labor is "human capital"; an idea that has survived to this day.

The early 20th century was characterized by an economic functioning that generated debates about whether the state should intervene in the economy. Production had ceased to depend solely on land, and the increasing amount and speed of movement of production due to industrialization had also begun to be experienced in the movement of money. According to classical economic theories, the economic order, which is known as a self-functioning mechanism and which favors the capitalist rather than the public, is also seen to have the ability to create crises. The economic crisis of 1929, known as the Great Depression, which could not be solved with existing approaches, convinced capitalists that it could only be solved by state intervention in the economy. John Maynard Keynes argued that the crisis was caused by a lack of demand and that it would be possible to restore the supply-demand balance by increasing public expenditures. Increasing public expenditures would be possible only, if the state, as an actor, took part in the market; that is, participated in the economy.

According to Keynes, who insisted that there was a permanent and stable relationship between unemployment and inflation, it should be the government, who could permanently reduce unemployment by increasing the demand for goods and services, although having to endure higher inflation. In the late 1960s however, Friedman challenged this view, arguing that if people were once again adjusted to a higher inflation rate, unemployment would climb again.

In a newspaper interview in 1970, Friedman argued that a 'businessman' who is self-elected or appointed directly or indirectly by shareholders, should be at once a legislator, an executive and a jurist. He or she would decide who should be taxed, how much, and for what purpose; and would herewith spend these taxes to rein in inflation, improve the environment, fight poverty, etc., guided only by general advice from on high (Friedman, 1970: 4). Considering Friedman's explanations for entrepreneurs, it is clear that the person who holds all this power, is of course a man, as the term 'businessman' implies. On the other hand, it also shows how patriarchal the economic system became in the 1970s, even as it developed. In this case, the man is the one who decides everything in important matters. What the man owns is important and valuable; he is the owner of both money and labor, and labor is seen as a power that earns money for its owner. In the patriarchal system, labor represents masculine power and is therefore considered very important.

The rapid increase in the world population after the 1800s, which was below one billion until the 1800s, has increased the search for more comprehensive and macro-scale validity in many issues in economic functioning, especially labor. As a continuation of the money-oriented culture of the capitalist system that took root in society along with the population, the understanding that started with adaptation in terms of expectations evolved into rationality in the New Classics. Every policy is ineffective, and practices that were found useful in the past are being renewed and tried to be implemented again. However, all developments in accordance with the nature of human development are reflected in interdisciplinary studies that require consideration of the diversity and unity among all human societies.

Although macro issues and their details are a separate research topic as well as the micro issues of economics, this study tries to theoretically calculate how domestic labor supplied in household work affects the labor factor as one of the components of GDP.

In historical progress, it is observed that labor has become a factor of production that is considered valuable and its price is determined by the capitalist and commodified depending on the condition of employment. The economic crisis, which occurs when the continuity of capital accumulation is disrupted and is characterized as a problem to be avoided and an undesirable process, is thought to create labor exploitation by making the system itself dependent on the use of more surplus value through employment. In other words, a crisis that only serves the continuity of capitalist production is defined, where it is claimed that the decrease in the amount of output produced depending on labor and capital inputs also negatively affects capital accumulation. The fact that the state and the civil society that constitutes the state have been left out of the game, has led to a process that has created many crises, not only economic but also social, environmental and cultural. The understanding of institutional governance, which has become widespread administratively and basically consists of the triad of state, capital and civil society, promises hope for economic management as well as public administration (Kumcu, 2009: 282).The social structure of the pre-capitalist period, in which the class differences to which the individual belonged were sharply demarcated, is being replaced by a human society that is transforming into an institutionalized society of equal individuals who learn lessons from the pains of today's rapid population growth.

In the transition from the labor-value relationship to the labor-benefit relationship, the determination of these relationships seem to have been overlooked. Considering that the raison d'être of the state, deep debates aside, is to pursue policies in favor of the citizen, the policies in favor of the owners of capital, who have become the determinants of economic functioning despite having no superiority over other actors, are becoming widespread. This is because the capitalist system has pushed the state out of the market by establishing the rule of "non-intervention of the state in the economy" since the beginning of its emergence. Thus, it seems as if the importance attached to the development of a system in which the "economic system" begins to exist has overridden the importance of the existence of the "state". What is overlooked at this point is that the owner of capital has replaced the state as the determinant of economic relations. This oversight becomes evident with the developments in governance models and is brought to the agenda again with the development of "governance"; an understanding of governance in which the triad of "state - civil society - capital" is balanced. To see the positive effects of governance in the economy, we can look at countries where economic governance models are applied. In undeveloped economies that try to be governed by adhering to the rule of non-involvement of the state in the economy, the governance model cannot be implemented, and even if it is implemented, it is not effective. Here, it is understood that respect for the labor of each individual and the equality of relations between the actors of the economy is a kind of governance responsibility.

While capitalism is rapidly globalizing with the phenomenon of self-creating crises, which ensures its continuity, it is seen that orthodox economic theories, which are known as mainstream and based on Neoclassical theories, cannot adapt to this speed to produce solutions to the crisis. The need for an interdisciplinary approach to economic thought by evaluating it together with all issues concerning human beings has brought heterodox approaches to the agenda.

Even if women have the same minimum income and have to meet the same living costs (house rent, food, clothing expenses, etc.), they are not considered to be fully compensated for their labor unless there is intra-family consultation as decision-makers or unless they participate equally with men in macro-scale law-making and administrative decisions.

It is thought that feminist economy is at the stage of criticizing the functioning, raising awareness and gaining acceptance within an existing capitalist economic philosophy with a patriarchal character; and that the mathematics of the theory, which is intended to include the female perspective, has not yet been revealed; since more work is needed on the econometrics of the motivated concept of equality. With this study, it-becomes evident that there is a need for the mathematics of feminist economy and the development of the theory by revealing other technical issues.

It can be seen that the paradigm shift from the rationality of the individual to limited rationality should be renewed in other micro and macro components within the science of economics. In the journey of human development, the guidance of science is needed at every step; and it is known that science is nourished through questioning and doubt. History shows that no system is more powerful, superior or indispensable than the rhythm of human development.

This paper argues about the concept of labor needing to be taken into account in the long run through the labor of housewives; because of their role in sustaining human capital.

In this study, the main objective contributing to labor theory, is to investigate whether the production of housework by a large portion of women, who constitute half of the population, has an impact on economic growth as a result of their preference to be housewives. Theoretical studies on female labor force show that an increase in female employment leads to an increase in economic growth. However, the positive effect of the increase in employment is not directly related to gender. In order to clearly calculate the contribution of housewives' labor, which is defined as women's labor to economic growth by evaluating their labor independently of the employment condition, we focus on the outputs they produce within the household.

The fact that women work both outside the home as much as to ensure equality in employment and inside the home, as much as not to interfere with housework and child development responsibilities shows the contradiction of GDP calculations, which attempt to make a calculation without double counting errors.

In the planning of economic growth, the efficient use and continuity of existing resources are considered. When growth models, that show how growth will occur for such planning or explain how it is realized are examined, it is seen that the parallelism between human capital owned by the labor force and production efficiency is preserved. In the case of housewives' labor, which is the focus of this study, this parallelism represents the condition of employment. It is seen that the condition for the inclusion of housewives in growth models is limited to their productivity in the workplace where they are employed, rather than their productivity at home.

In this study, a humanist model is proposed to question whether housewives' labor is an economic asset. The model is also important in that it contains findings that show how the economy is at full "employment", or rather, how labor is independent of employment. In the calculations made with the proposed model, it is shown that the LGSDP, which is the GDP that takes into account the production of housewives' labor within the household, is larger than the GDP that does not take into account housewives' labor. Inequality coefficients are calculated for the two cases where housewives' labor is not taken into account and for the two cases where housewives' labor is taken into account; and it is shown that by taking housewives' labor into account, inequality is eliminated and a more equal distribution is possible. This finding, which shows the degree of welfare created by the presence of housewives' labor in this distribution, is expected to shed light on the explanation of unemployment growth.

In growth models that consider the female labor force as wage labor, where women are employed as waged but cheap labor, productivity is calculated according to the skilledness of labor and women are characterized as low-productivity and unskilled workers due to reasons such as lack of education and limited working hours. Unlike wage-earners, no economic growth model provides any compensation for the labor of housewives. The way to achieve equality in this situation would be to recognize mutual interests and relationships.

The paradigm on labor continues to evolve. Philosophical and logical clarity is needed to mathematically calculate the labor of housewives. The accumulation of health, education and social sciences on human conception and healthy growth also shows that the indispensability of the family for the continuation of the human race is a reality that has persisted from past to future. If it were not for genetic transmission, maternal education would be much more costly and long-termed. For this reason, globalizing world economies are expected to adopt policies to ensure that the labor of housewives, which is the accumulation of being a woman, is included in production and distribution as an economic value, even if she is only a mother.

It is predicted that the number of limited rational individuals will increase in parallel with the increasing use of technology in production models and will not only consist of housewives. Examples of technology changing the types of energy used in production are evident in the shift from labor power to machine power. However, it is expected that the uncertainty of where human energy can be used will become more understandable and explainable as applications in the virtual world, the so-called metaverse, become more widespread. Therefore, the paradigm of defining and interpreting labor in relation to energy in economic theories is one of the expected directions for future work.

In a world with very "traditionalist" views on the question of whether a job is best for economic independence or not, it is also true that the fulfillment of household chores is important. It is also true of economic theories, that the answer to this question, seemingly traditional, might have changed over time. The notion that housework should be seen as something to be contended with, could be seen as a realistic and humane 'reinterpretation of housewives' labor' rather than a traditional view that reflects the idea that women's place is the home.